\documentclass[preprint,prb,showpacs,showkeys]{revtex4-1}
\usepackage{amsfonts}
\usepackage{amsmath}
\usepackage{amssymb}
\usepackage{graphicx}
\usepackage{subfigure}

\begin{document}

\title{High-pressure behavior of dense hydrogen up to {3.5\,TPa} from density functional theory calculations}
\author{Hua Y. Geng}
\affiliation{National Key Laboratory of Shock Wave and Detonation
Physics, Institute of Fluid Physics, CAEP;
P.O.Box 919-102 Mianyang, Sichuan, P. R. China, 621900}

\author{Hong X. Song}
\affiliation{National Key Laboratory of Shock Wave and Detonation
Physics, Institute of Fluid Physics, CAEP;
P.O.Box 919-102 Mianyang, Sichuan, P. R. China, 621900}

\author{J. F. Li}
\affiliation{National Key Laboratory of Shock Wave and Detonation
Physics, Institute of Fluid Physics, CAEP;
P.O.Box 919-102 Mianyang, Sichuan, P. R. China, 621900}

\author{Q. Wu}
\affiliation{National Key Laboratory of Shock Wave and Detonation
Physics, Institute of Fluid Physics, CAEP;
P.O.Box 919-102 Mianyang, Sichuan, P. R. China, 621900}


\begin{abstract}
Structural behavior and equation of state of atomic and molecular crystal phases
of dense hydrogen at pressures up to 3.5\,TPa
are systematically investigated with density functional theory.
The results indicate that the Vinet EOS model that fitted to low-pressure experimental data
overestimates the compressibility of dense hydrogen drastically when beyond 500\,GPa.
Metastable multi-atomic molecular phases with weak covalent bonds are
observed. When compressed beyond about 2.8\,TPa, these exotic low-coordinated phases become competitive with the groundstate
and other high-symmetry atomic phases.
Using nudged elastic band method, the transition path and the associated energy barrier between these high-pressure phases
are evaluated.
In particular for the case of dissociation of diatomic molecular phase into the atomic metallic Cs-IV phase,
the existent barrier might raise the transition pressure about 200\,GPa at low temperatures.
Plenty of flat and broad basins on the energy surface of dense hydrogen have been discovered, which
should take a major responsibility for the highly anharmonic zero point vibrations of the lattice, as well as
the quantum structure fluctuations in some extreme cases.
At zero pressure,
our analysis demonstrates that all of these atomic phases of dense hydrogen known so far are unquenchable.

\end{abstract}

\volumeyear{year}
\volumenumber{number}
\issuenumber{number}
\eid{identifier}
\maketitle


\section{INTRODUCTION}
Having only a single electron outside the nucleus, hydrogen is the simplest and most abundant element in the universe.
It is also an essential element for models of stellar and planetary interiors.\cite{guillot99a,guillot99b}
Hydrogen shows characteristics of both the group \uppercase\expandafter{\romannumeral 1}
alkalis and the the group \uppercase\expandafter{\romannumeral 7} halogens. At low pressures,
hydrogen isotopes are halogenous, covalent diatomic molecules that form insulators.
Yet at high pressures, it is one of the most difficult to understand.
It displays anomalous melting behavior with a maximum in the melting temperature versus pressure curve
at high temperatures,\cite{bonev04,deemyad08} and undergoes a first-order liquid-liquid transition
under further compression.\cite{morales10pnas,lorenzen10}
At low temperatures, it is
experimentally known that hydrogen can exist as a rotational crystal (phase \uppercase\expandafter{\romannumeral 1})
on a hexagonal close-packed (HCP) lattice to
high pressures ($P<110$\,GPa), followed by a transition into the broken-symmetry phase \uppercase\expandafter{\romannumeral 2}
($110\,\text{GPa}<P<150\,\text{GPa}$) which is marked by a change to wide-angle libration and hence to a continuing
incoherence of motion between different molecules,
and then to phase \uppercase\expandafter{\romannumeral 3} at about 150\,GPa.\cite{mao94}
Possible pressure-induced insulator-metal transition also has been extensively studied up to 320\,GPa.\cite{mao94,eggert91,narayana98,loubeyre02}
However, beyond the fact that protons remain paired within this
pressure range, their time-average locations are to date experimentally unknown,
mainly because hydrogen atoms scatter X-rays only weakly, leading to low-resolution diffraction patterns.\cite{mao94,bafile08}
Experimental data at higher pressures are scarce, and insightful understanding of structural behavior of ultra-dense hydrogen is lacking.

Theoretical prediction of stable crystalline structures and properties of dense hydrogen at
high pressures has been pursued for decades.\cite{wigner35,friedli77,barbee89,chacham91,johnson00,pickard07,tse08,mcmahon11}
It is difficult because of the need to search the very large
space of possible structures, and the necessity of obtaining accurate energies for each of these structures.\cite{pickard07,mcmahon11} First-principles
density functional theory (DFT) has been proved as an efficient approach of calculating quite accurate energies, and has provided
insights into properties of various materials, including solid hydrogen under compressions. At present, DFT offers a high level
of theoretical description at which we can carry out searches over many possible candidate structures.\cite{johnson00,pickard07,tse08,mcmahon11} Recent
DFT calculations have predicted that within the static-lattice approximation, the most stable phases of dense hydrogen are $P6_{3}/m$
($<$105\,GPa), $C2/c$ (105-270\,GPa), $Cmca$-12 (270-385\,GPa), and $Cmca$ (385-490\,GPa), followed by atomic $I4_{1}/amd$ (Cs-IV)
phase,\cite{pickard07,tse08} and then $R3m$ (or $R\overline{3}m$).\cite{mcmahon11}

On the other hand, at pressures high enough so that electrons are fully ionized out to form a uniform background,
hydrogen becomes one-component plasma.\cite{wigner34} In this regime the dominant
Coulomb interaction forces the ions into an ordered configuration called Wigner crystal, and
stabilizes in a body-centered cubic (BCC) structure.\cite{wigner34,carr61,jones96}
Except this, the general structural and compressional behavior of dense hydrogen at ultra-high pressures
beyond several Mbar (1\,Mbar=100\,GPa) are still poorly understood,
in spite of a few theoretical investigations available.\cite{pickard07,mcmahon11}
In particular, the energy barriers of pressure-induced phase transitions are completely unknown, especially that
of molecule dissociation. A high energy barrier would lead to a hysteresis and
raise the transition pressure from what it otherwise would be.\cite{geng07}
Another important issue that has not been solved completely is about the applicability of extrapolating the equation of state (EOS) that fitted at low pressure to higher pressures.
This is an intriguing problem because we knew that the EOS model fitted
to lower-pressure data (up to 42\,GPa)\cite{mao94} fails to capture (underestimates) the compressibility of
H$_{2}$ and D$_{2}$ at the relatively higher pressures,\cite{loubeyre96} which is due to the transition of H$_{2}$ (or D$_{2}$)
from a freely rotating phase into a broken-symmetry phase. At ultra-high pressures, diatomic hydrogen dissociates
into atomic phase, and this transition might invalidate the early established EOS model again very likely.

The groundstate structures of atomic metallic hydrogen from 500\,GPa to 4.5\,TPa
has been extensively searched by McMahon \emph{et al.}\cite{mcmahon11} using
the \emph{ab initio} random structure searching (AIRSS) method.\cite{pickard06}
Their work was carried out for unit cells containing only 4 and 6 atoms.
However, as recent investigations suggested, complex structures can be adopted by simple alkali metals
at high pressures.\cite{rousseau11,lv11} Similar phenomenon might also occur in dense hydrogen.
Furthermore, there is no absolute guarantee for a specific structure searching approach
to find the true ground state within a finite computational time due to the limited phase space it can explore.
There are examples that AIRSS calculations failed to detect lower energy states which later were captured
by other structure searching method such as evolutionary algorithms.\cite{yao09,lv11}
Therefore a cross-check of the proposed groundstate structures with alternative methods
is always necessary to ascertain the results.
With systematic and accurate first-principles calculations using evolutionary searching
method combined
with particle swarm optimization algorithm,\cite{wang10} we confirm in this work that
Cs-IV atomic phase with a space group of $I4_{1}/amd$ becomes
the most stable phase beyond 490\,GPa, but a novel structure $Fddd$ is also found
to be \emph{degenerate} with it over a wide range of pressure.
These two metallic phases persist up to 2.3\,TPa, where exotic multi-atomic structures with
low symmetry becomes competitive. Our calculation also indicates that dense hydrogen
has many very broad and flat basins on the energy surface, which
has never been noticed before. This flatness of the energy variation not only leads to great anharmonicity in zero-point vibrations of lattice,
but also blurs the \emph{boundary} of some {crystal structures}.
Furthermore, the transition paths between high pressure phases are modeled with nudged elastic band (NEB)
method, and the associated enthalpy barriers and energy variations are analyzed.
In the next section we will present the methods for total energy and NEB calculations.
The results and discussion are given in Sec. \ref{sec:result}, followed by a summary in Sec. \ref{sec:sum}.

\section{Methodology}
\subsection{Total energy calculation}
The total energy calculations were performed with DFT as implemented
in the Vienna Ab-initio Simulation Package (VASP).\cite{kresse96}
The electronic structure was described with
all-electron like projector augmented-wave (PAW) potential,\cite{blochl94,kresse99} and the
Perdew-Burke-Ernzerhof (PBE) exchange-correlation
functional was used.\cite{pbe96}
A hard version of the PAW potential that is specially designed for high pressure applications
was employed.
The k-point sets were generated separately for each unit cell encountered during the procedure,
and a high quality Brillouin zone sampling with
a grid of $15\times15\times15$ were found to be sufficient for structure optimization.
When re-calculate the enthalpy curves, we used a denser k-point mesh that can generate at least 1500 irreducible
points. The residual Pulay stress was removed by increasing the kinetic energy cutoff of the plane wave basis set to 900\,eV,
which was confirmed by observing the vanishment of the difference between the Hellmann-Feynman pressure and
that computed from the energy curve with $P=-dE/dV$. Increase the energy cutoff to 1200\,eV led to a pressure
change less than 0.1\,GPa. A justification of the methods,
especially DFT and the employed pseudopotential, is given in appendix.
With this parameter setting, the structural features are estimated to be converged to better than 0.2\%,
the convergence of total energy is to within 3\,\emph{m}eV per proton,
and the relative energy
difference between structures is to within 1\,\emph{m}eV per proton.

By minimizing the enthalpy, we carried out extensive and systematic structure searches of molecular and atomic
phases at pressures up to 3.5\,TPa.
All structures were fully relaxed at fixed volumes with a force tolerance of 0.1\,\emph{m}eV\AA$^{-1}$.
The pressure was directly computed with the Hellmann-Feynman theorem, which then was used to calculate the enthalpy.
In addition to diatomic molecular phases,
many high-pressure candidates of atomic phase were considered and their enthalpies were calculated, including
cubic structures (SC, BCC, and FCC) and their low-symmetry distortions ($\beta$-Po, $\beta$-Hg, In-{\uppercase\expandafter{\romannumeral 1}},
and In-{\uppercase\expandafter{\romannumeral 2}}), hexagonal structures (HCP and $\omega$-Ti), diamond structure,
$I4_{1}/amd$ ($\beta$-Sn and Cs-{\uppercase\expandafter{\romannumeral 4}}) phases, and so on.\cite{mcmahon06}
Details are presented in the following subsections.

\subsection{Structure search}
The search for high pressure structures of dense hydrogen was mainly
carried out with the particle swarm optimization (PSO) technique\cite{wang10} within an evolutionary scheme that combined
with first-principles total energy calculations using VASP.
PSO is an efficient approach of evolutionary methodology but quite different from genetic
algorithm (GA). In particular the major evolution operations of crossover and mutation in GA have been
avoided. PSO has been verified to perform well on many optimization problems.\cite{wang10,lv11,rousseau11}
In this work, stable and meta-stable structures containing up to 24 atoms per unit cell at pressures up to 3.5\,TPa
were automatically explored and generated by CALYPSO code,\cite{wang10} which implements PSO algorithm.
With a population size of 30 in each generation and a total allowed number of generations of 30,
it is sufficient to ensure the convergence of the structure search.

As a complement to the CALYPSO search, in order to better understand the relationship between
symmetry and the structure stability, as well as to track the evolution of structures with pressure, we also performed
local structure searches manually.
It was based on the already
known information about stable and meta-stable phases at relatively low pressures.\cite{johnson00,pickard07}
Some molecular and
atomic phases which were not yet fully investigated previously were added into the
structure library of search. The atomic phases were selected from Ref.\onlinecite{mcmahon06} by
consulting previous theoretical predictions to pick out the most likely candidates, while putting
high weight on low-symmetric structures.

The initial structures for optimization were generated from the search library by drifting the ionic positions
and distorting the lattice shape randomly. Interpolated configurations between the most stable phases
were also taken into account.
Although most of the phases in the search library
are locally stable, with large enough distortions new
structures can always be produced.
To eliminate a possible sensitive dependence of the final results
on the initial configurations, the ionic relaxation procedure employed algorithms of both quasi-Newton
and conjugate-gradient method alternatively.
The derived structures were then fully relaxed to
the energy minima at constant volumes without any symmetry restrictions.
After the ionic optimization process converged, we calculated the pressure and then the
corresponding enthalpy. At each pressure, the search process was repeated several times (each
iteration involves all structures in the search library
that was updated accordingly) to confirm the results, and in total 4 typical
pressure values (around 0.5, 1, 1.5 and 2\,TPa, respectively) were explored.
The whole enthalpy curves of the low-lying phases were then computed.

\subsection{NEB calculation}
Structural phase transition paths and the associated energy barriers were modeled using the nudged elastic band (NEB) method\cite{mills95}
as implemented in the VASP code, which searches for the minimum energy path (MEP) by moving a chain of ionic configurations
or images that bridges the initial and the target structures. Tangential springs were introduced to keep the
images being equidistant during the relaxation. The potential energy maximum along the MEP is the
saddle point energy which gives the activation energy barrier.

In calculations, the supercell method was employed, which in most cases contains 12 atoms, and in some special situations
a cell with 24 atoms was also used. To model the transition path with a variable cell volume and
shape using NEB technique, the continuous MEP was constructed by linking the images with springs only in a
pre-aligned supercell (\emph{i.e.}, the images due to periodic boundary conditions were discarded).
When computing the transition path from high-pressure dense phases to
a diatomic molecular phase at 0\,GPa,
the molecular structure was generated by distorting and then relaxing the corresponding dense hydrogen phase respectively.
This strategy simplifies the NEB calculations greatly since in this way the initial path generated through
a linear interpolation of the initial
and the target configurations gives a good approximation to the final transition path.
Then a minimization of the whole system was carried out to trace out the MEP.
In all calculations at least five images (seven if includes the two end images) were employed, which in most cases is sufficient to give an
acceptable resolution to the discrete representation of the MEP.

\subsection{Calculation of phonons in harmonic approximation}

Zero-point (ZP) vibrations of protons were neglected during the structure search and optimization procedure,
but with a subsequent estimation of its impact on the relative stabilities using harmonic
ZP energy (ZPE) calculations. The full phonon spectra in a harmonic approximation were calculated with the small-displacement
method as implemented in the PHON code.\cite{phon} Big enough supercells containing
more than 64 atoms were used. In the associated DFT calculations, a Brillouin zone sampling mesh of
$20\times20\times20$, a kinetic energy cutoff of 1000\,eV, and a very dense support augmentation
charge grid that is required for an accurate force calculation were used.
This setup gives a convergence in the ZPE better than 2\,\emph{m}eV per proton.
The magnitude of the small displacement
was slightly varied to check the numerical stability of the calculated force constant matrices. The
ZPE was estimated from the phonon density of states $g(\omega)$ by $\int g(\omega)\hbar\omega/2\;d\omega$,
and the ZP mean square displacement (MSD) was calculated by
$\int g(\omega)\hbar/(2M\omega)\;d\omega$, where $M$ is the mass of hydrogen atom.

\section{results and discussion}
\label{sec:result}
\subsection{Static structure and enthalpy}
\subsubsection{Ground-state of dense hydrogen}
\begin{figure}
  \includegraphics*[width=3.0 in]{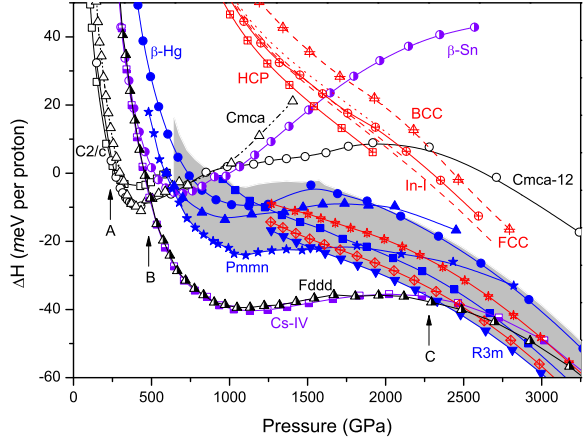}
  \caption{(color online) Enthalpy difference per proton as a function of pressure with respect to a reference state.
  Arrows indicate the main phase transitions of the ground state and the shadow marks the region where the
  meta-stable low-symmetry multi-atomic phases lie in.
  Typical curves are labeled as follows: open square--$C2/c$,
  open circle--$Cmca$-12, open triangle--$Cmca$, filled star--$Pmmn$, filled square--$C2/m(2)$, filled up-triangle--$C2/m(1)$,
  filled circle--$\beta$-Hg, half-filled square--Cs-IV, half-filled triangle--$Fddd$, filled down-triangle--$R3m$, crossed rhombus--$R\overline{3}m$, crossed pentagram--$P6_{3}/mmc$,
  crossed triangle--BCC, crossed circle--FCC, and crossed square--HCP.
  }
  \label{fig:dH}
\end{figure}

The calculated enthalpy difference with respect to a reference state (virtually defined by its enthalpy as $H=-3.985+0.216P^{0.55}$\,eV/proton, $P$ in a unit of GPa)\cite{note_ref}
of the most stable phases are shown in Fig.\ref{fig:dH}. Overall, our results agree very well with
previous theoretical predications: diatomic $C2/c$ is the groundstate at 110\,GPa and transforms into $Cmca$-12 at about
255\,GPa;\cite{pickard07,tse08} at 370\,GPa, $Cmca$ overtakes slightly,\cite{pickard07} but it soon degenerates into $Cmca$-12,
which is stable until
an atomic phase-- Cs-IV with space group $I4_{1}/amd$\cite{johnson00} and its distortion $Fddd$-- becomes the groundstate at 495\,GPa;
beyond 2.3\,TPa, the most favored phase is shifted to trigonal $R3m$ in a static lattice approximation.\cite{mcmahon11}

The main transition points are indicated by arrows in Fig.\ref{fig:dH}.
Point B is an intersection of the molecular phases and the low-lying
atomic Cs-IV and $Fddd$ phases. Namely, it is a pressure-induced dissociation point.
This is in agreement with previous DFT calculations,\cite{johnson00,pickard07,mcmahon11} except for a newly discovered degenerate phase
$Fddd$ that was not detected before. $Fddd$ is an orthorhombic variant of the diamond phase,
but is also close to being a distortion of Cs-IV locally.
Beyond the transition point C, there are many structures with closely competitive enthalpy,
reflecting the frustration among competing factors. In this regime, a small change
in interactions will tip the balance from favoring one phase into another. Although $R3m$ seems to have the lowest static lattice enthalpy,
harmonic phonon calculations indicated that the zero point vibrations of lattice might make
$R\overline{3}m$ be more favored.\cite{mcmahon11} It is interesting to note that both Cs-IV and $Fddd$ have 4
nearest neighbor (NN) atoms, and $R\overline{3}m$ has 6 as its first coordination number (CN).
In contrast, $R3m$ apparently has just 2 NNs, but since the distance from its first NN shell to the next one is very short
(0.06\,{\AA} at 3\,TPa), so that it in fact has 6 atoms in its first coordination shell.
From this point of view, the evolution of the groundstate structures of dense hydrogen under compression
is evident: with increasing density, the structure evolves from with 1 NN (diatomic molecular phases)
to that with 4 NNs (atomic Cs-IV and $Fddd$), and then to that with 6 NNs ($R3m$ or $R\overline{3}m$), finally transforms to
cubic structures with 8 (BCC) or 12 (FCC) NNs.\cite{mcmahon11}

\begin{figure}
\centering
\begin{tabular}{r l c r}
\subfigure[]{\label{fig:struct:a}\includegraphics[width=1.45 in]{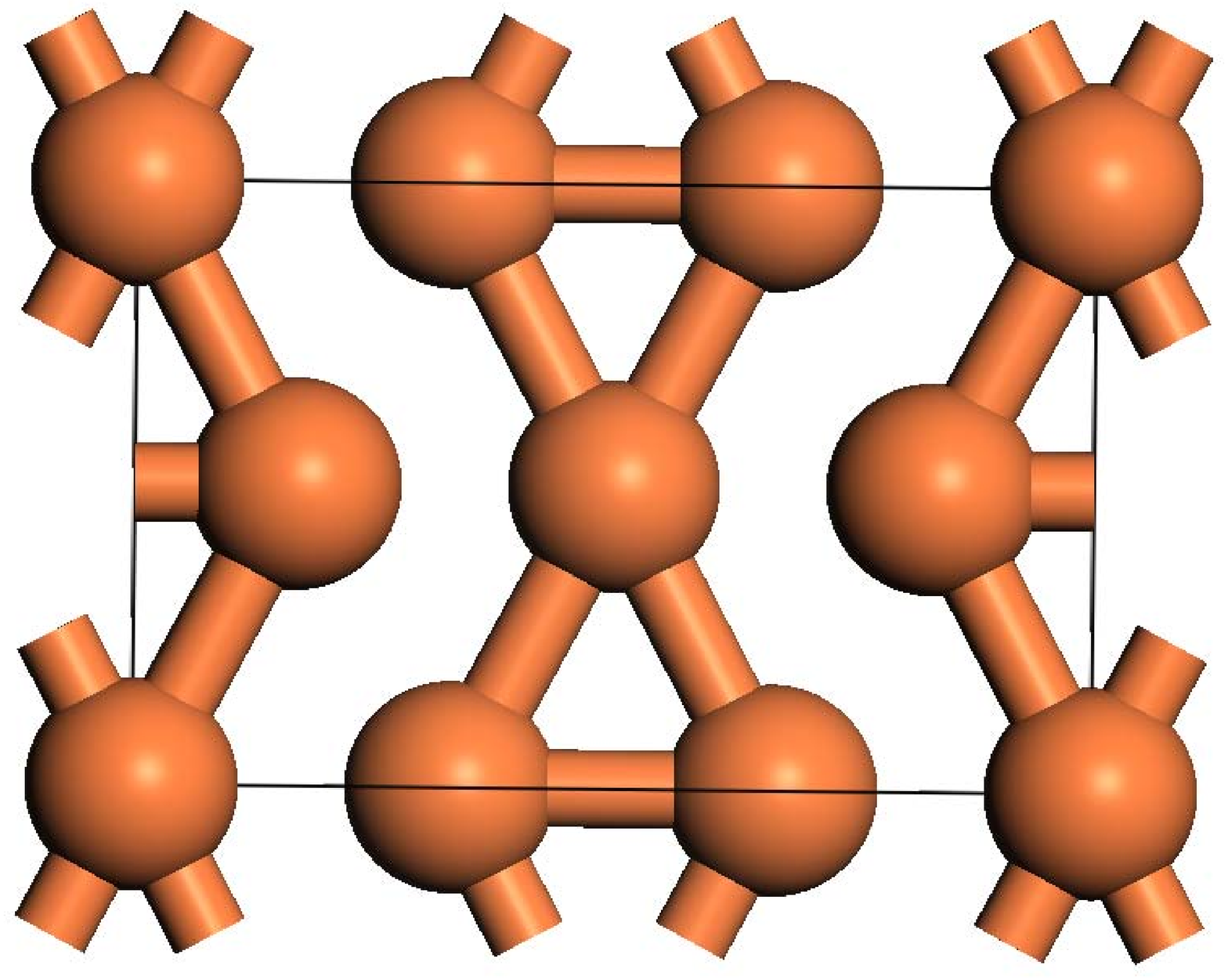}}
\hfil &
\subfigure[]{\label{fig:struct:b}\includegraphics[width=1.4 in]{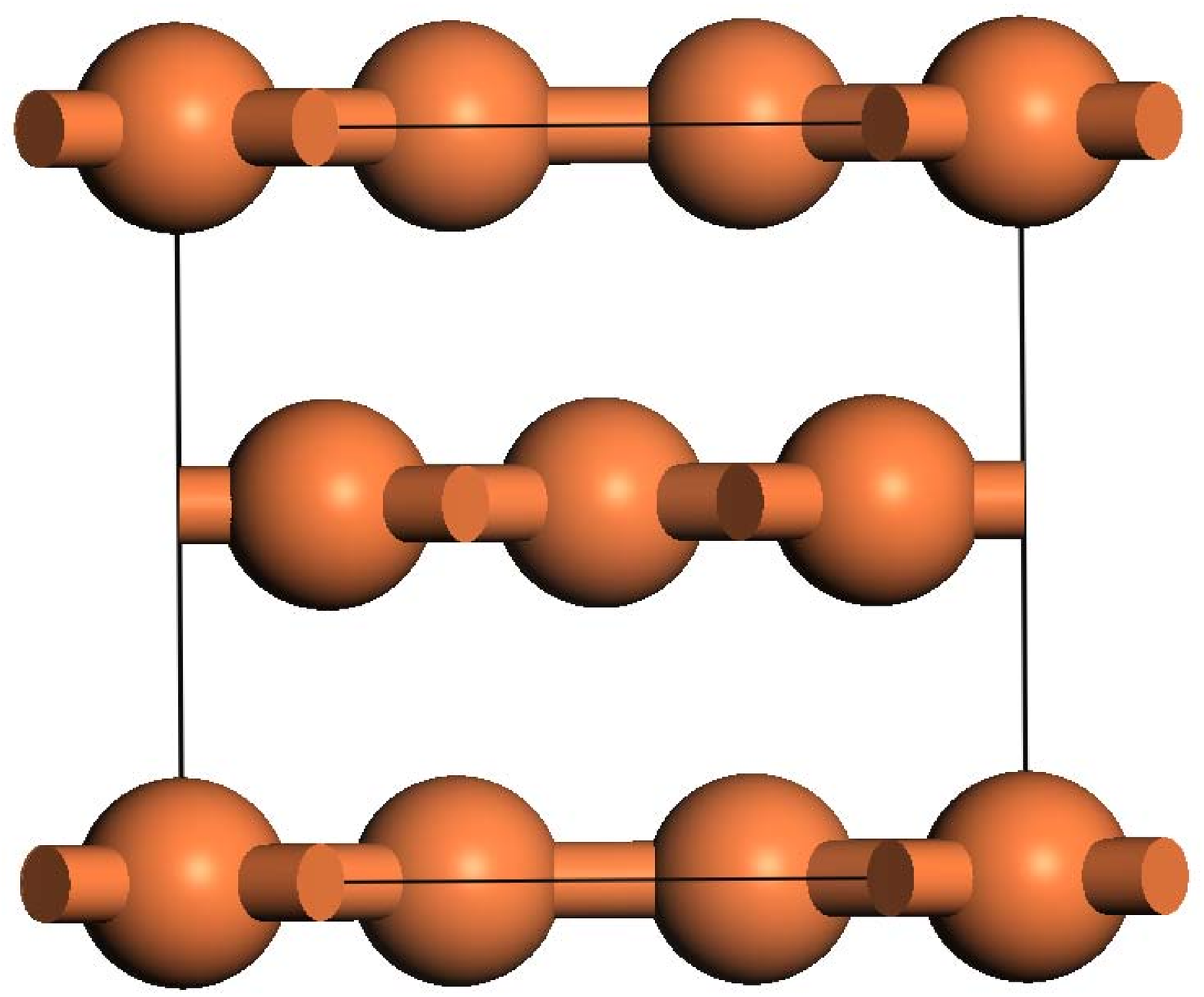}} &
\hfil &
\subfigure[]{\label{fig:struct:c}\includegraphics[width=1.05 in]{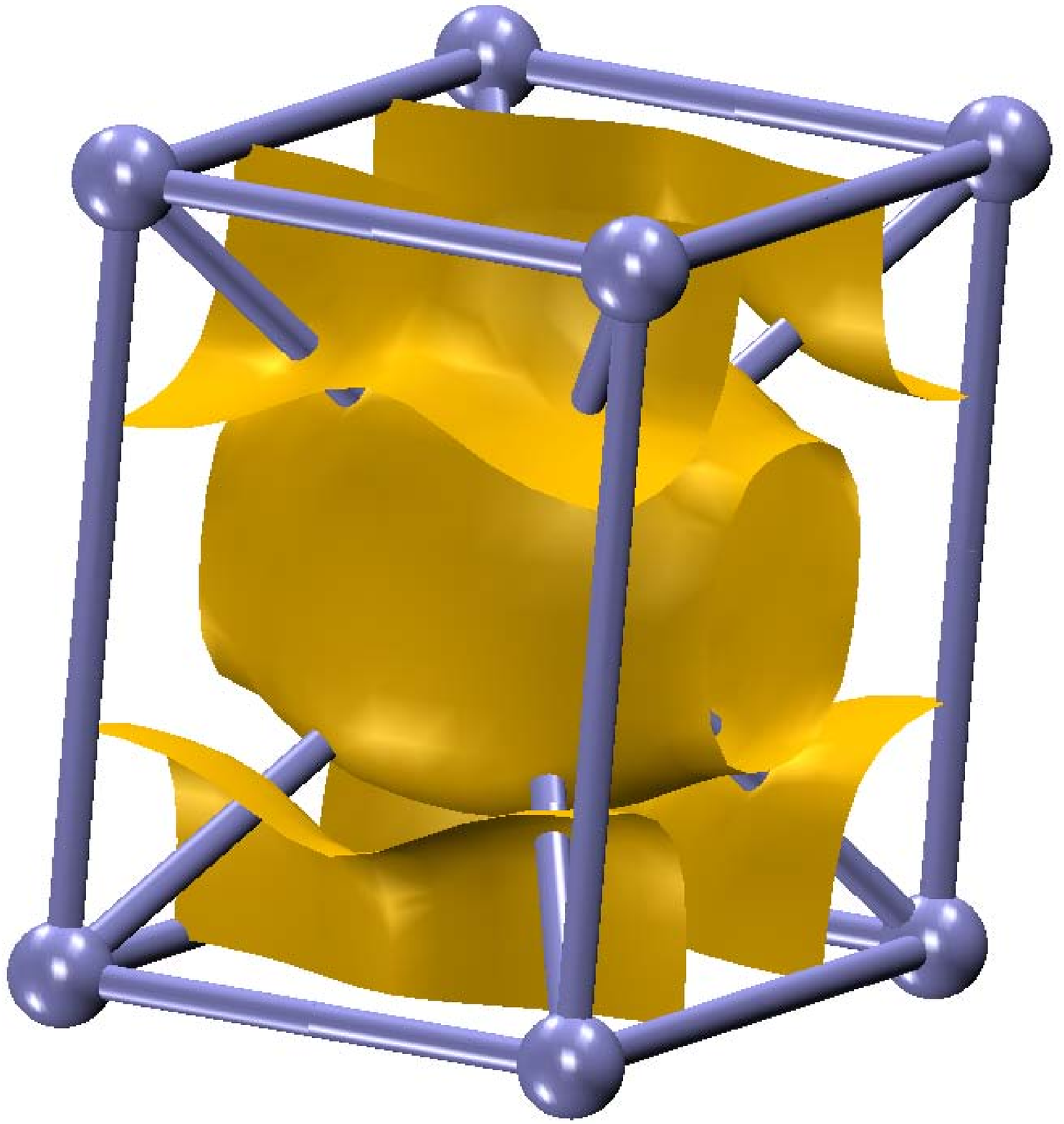}}
\end{tabular}
  \caption{(color online) Structure of (a) $C2/m(2)$, (b) $C2/m(1)$, and (c) $\beta$-Hg. A charge density isosurface
  also illustrates in (c).
  The two $C2/m$ phases (both in $Immm$ representation and projected onto the same plane)
  are different only in the connection manner of the covalent bonds, and $\beta$-Hg is derived from
  BCC structure with covalent bonds formed along the shortest $c$ direction.}
  \label{fig:struct}
\end{figure}

\subsubsection{Metastable multi-atomic molecular crystals}

Other monatomic phases with high symmetry, such as BCC, FCC, and HCP, and their
distortions, do not approach the ground-state line before 3.5\,TPa
(the simple cubic (SC) phase is always 50\,\emph{m}eV higher than others and not shown here).
One interesting phase is $\beta$-Hg, which
caught no attention before. Its enthalpy is already low enough at 600\,GPa as shown in Fig.\ref{fig:dH}, and approaches gradually to
the groundstate line all the way beyond 3\,TPa. It might coexist with other meta-stable phases such as $Pmmn$ and $C2/m$
over a wide range of pressure. In conventional sense it should be a monatomic phase distorted from BCC structure
with a shorter lattice length in the $c$ direction.\cite{mcmahon06} But for hydrogen at pressures beyond 600\,GPa this lattice length is already short
enough so that covalent bonds are established along this direction and it becomes
a chained molecular state (see Fig.\ref{fig:struct:c}). In fact it is exactly the formation of this kind of multi-atomic molecular bonds
that stabilizes the structure, due to a combined effect of
pressure induced excitation of the $1s$ electrons (which weakens the diatomic covalent bond)
and the tendency to overlap electronic orbitals among neighboring atoms. It can be seen much clearly by comparing the enthalpy with that of
$Cmca$ (diatomic bonds), $Pmmn$ (triatomic bonds, see below), and $C2/m$ (molecular chains).
Specifically, $\beta$-Hg constitutes linear chains of H$_{2}$ bonds, whereas
$C2/m$ consists of chains of H$_{3}$ clusters. $C2/m(1)$ and $C2/m(2)$ belong to the same space
group. The only difference is in the linkage pattern of the bonds and the chain orientation. In addition, $C2/m(1)$
has a shorter interchain distance and trends to form a bond network, whereas $C2/m(2)$
has more localized bonds and thus its structure is much distinct and well defined.
Figures \ref{fig:struct}$\sim$\ref{fig:H-Pmmn} and table \ref{tab:c2m2} in appendix provide detailed structural information about these phases.
It is worthwhile to point out that the $C2/m$ phase reported in Ref.\onlinecite{mcmahon11} corresponds to the $C2/m(1)$ here,
and $C2/m(2)$ has a lower enthalpy at high pressures. The existence of this isomorph (and others alike) indicates
the complex nature of the structure of dense hydrogen.

\begin{figure}
\centering
\begin{tabular}{r l}
\subfigure[]{\label{fig:pmmn:a}\includegraphics[width=2.5 in]{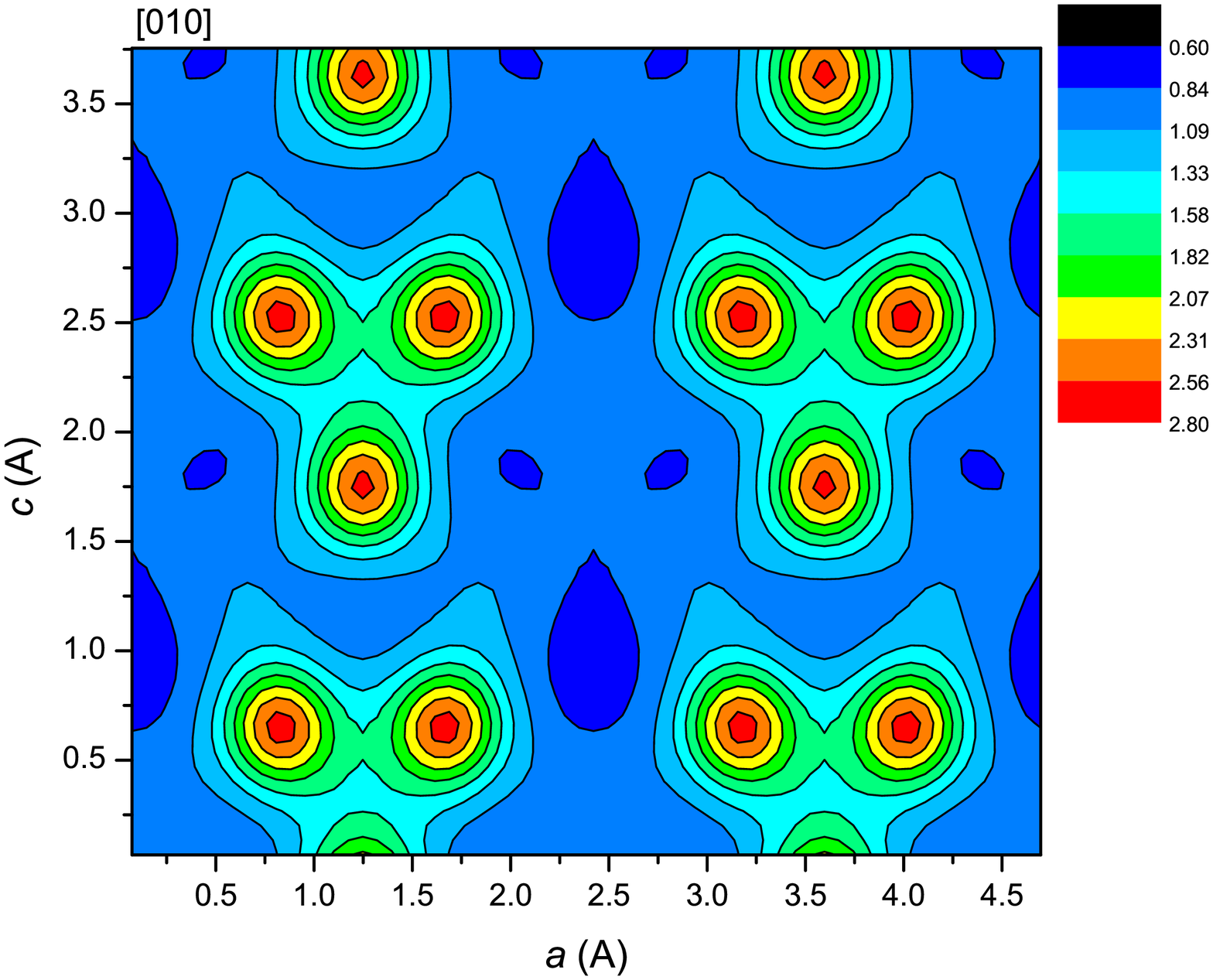}}
\hfil &
\subfigure[]{\label{fig:pmmn:b}\includegraphics[width=2.5 in]{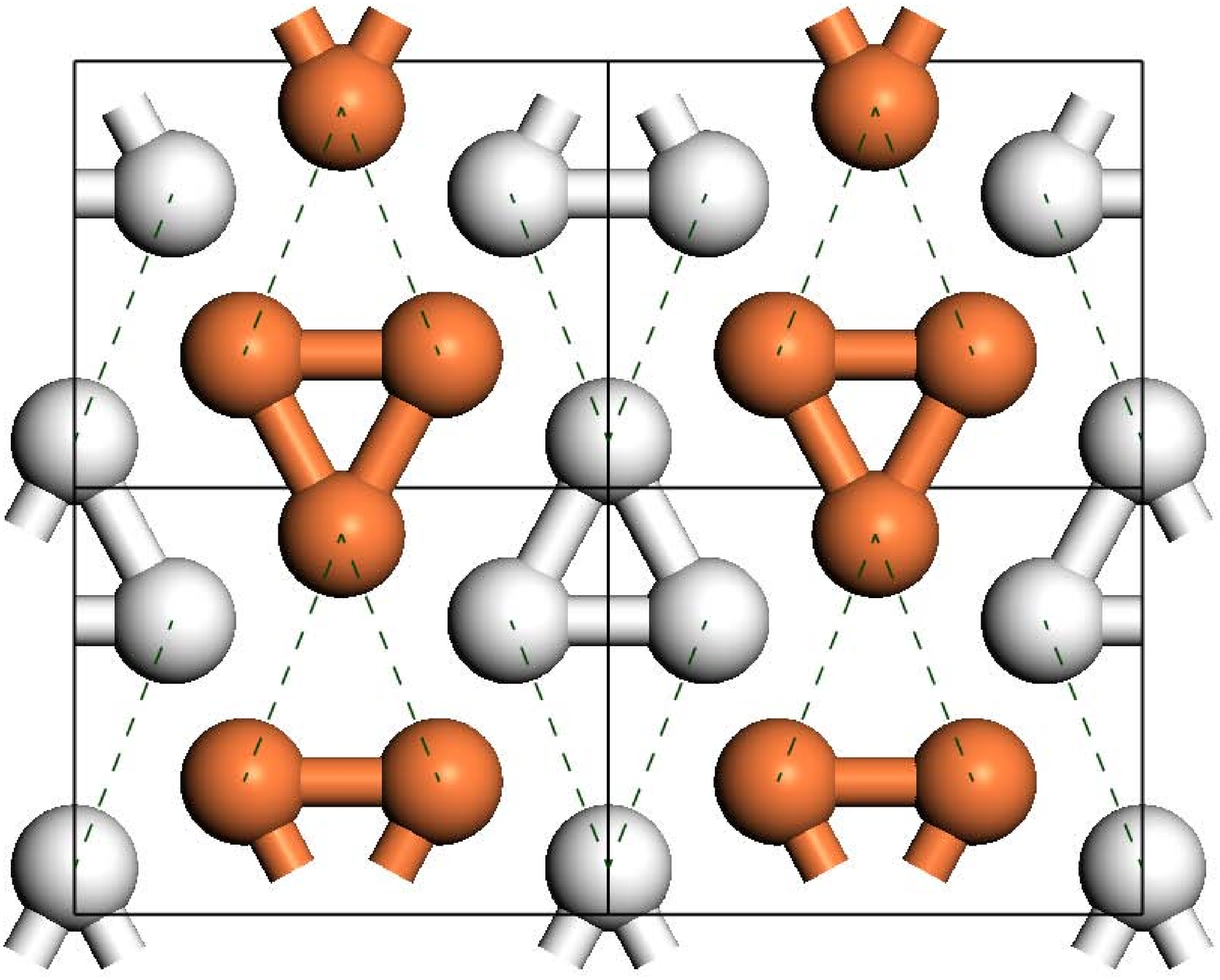}}
\end{tabular}
  \caption{(color online) Charge density on [010] plane (a) and the atomic structure (b) of
  a 2$\times$1$\times$2 supercell of $Pmmn$ phase at 1104\,GPa,
  in which the H$_{3}$ clusters are evident. In (b): atoms on alternating layers are distinguished by
  different color/grey scale, and the dashed lines indicate the connection manner of covalent bonds in a chained molecular phase
  $C2/m(2)$.}
  \label{fig:H-Pmmn}
\end{figure}

Although $Pmmn$ is a triatomic phase (see Fig.\ref{fig:H-Pmmn}) and $C2/m(2)$ is a chained molecular structure (see Fig.\ref{fig:struct:a}),
they are closely related.
Both phases can be drawn on an orthorhombic lattice, and
their structural relationship manifests clearly in
an $Immm$ representation. By comparing Figs. \ref{fig:pmmn:b} and \ref{fig:struct:a} (and also their charge densities),
it is easy to find the similarity of these two structures: by compressing the former phase along its $c$ direction to
create new bonds along the dashed lines as indicated in figure \ref{fig:pmmn:b} meanwhile shifting the relative position of the alternating layers slightly
along this direction, one gets the latter phase. This is actually
how the structural transition from $Pmmn$ to $C2/m(2)$ takes place at 1.9\,TPa.
It is reasonable
since compression reduces inter-distance among H$_{3}$ clusters, and thus could establish chains of covalent bonds
along certain direction if the strain tensor is anisotropic.

It is worthwhile to note that at the intermediate pressure range the NN number of $C2/m$ phases is about 3, and both $Pmmn$
and $\beta$-Hg have a NN number of 2. They all are multi-atomic molecular phases. Furthermore,
distorting some simple high-symmetric structures can make them continuously relax to
configurations with a low CN, and the resultant enthalpy lies in the shaded range as shown in Fig.\ref{fig:dH}.
The above mentioned multi-atomic molecular phases are members of these low-symmetric structures.
This connection provides a plausible answer to the questions of why structures with a CN of 2 or 3
absent from the groundstates, and why dense hydrogen dissociates into orthorhombic atomic Cs-IV
instead of simple high-symmetric monatomic phases such as BCC, FCC, or HCP.
As it is well known, hydrogen has just one
electron outside the nucleus. This makes the electron cloud be compact and tightly
attached to the nuclei even at high pressures.
On the other hand, at a given density, high-symmetric monatomic phases always have a greater interatomic
distance, making it difficult for hydrogen to share electron with its neighbors.
In this sense, distortion of the structure to reduce interatomic distance between some atoms can effectively facilitate
wavefunction overlapping and lower the energy.
That is the reason why dense hydrogen does not dissociate into
simple monatomic phases directly.
From another point of view, diatomic covalent bond has been weakened greatly
by compression at high pressures, and thus prevents a further transition of these low-symmetric phases into
a diatomic molecular configuration via mechanism-- for example, the Peierls instability\cite{peierls55}-- from happening when the pressure is higher than 500\,GPa.
However, because each hydrogen has only
one electron, it is not easy to form stable multi-atomic (triatomic or molecular chain) bonds.
Even with the aid of a small portion of $p$ electrons that are excited from $s$ orbitals by compression
(this increases the anisotropy of the electron cloud and
thus favors multi-atomic bonds),
the multi-atomic molecular phases
are just be meta-stable, and the balance can be easily tipped down towards the atomic phases with
a low CN, which takes advantages of all competing factors.
That explains why no groundstate of dense hydrogen has a NN number of 2 or 3.

\begin{figure}
  \includegraphics*[width=3.0 in]{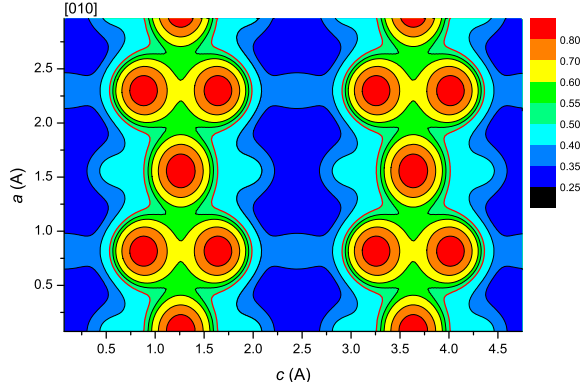}
  \caption{(color online) Electron localization function of the $C2/m(2)$ phase (in a tetragonal $Immm$ representation) at about 2.7\,TPa.
  Formation of the molecular chains is evident. }
  \label{fig:c2m2}
\end{figure}

\subsubsection{Electronic structure}

Figure \ref{fig:pmmn:a} shows the calculated charge density of $Pmmn$
at about 1\,TPa.
The tri-atomic molecules are evident. Note that the environmental
charge density is nonzero and the multi-atomic molecules are in fact submerging in a sea of electrons that mediates
metallic interactions.
Figure \ref{fig:c2m2} plots the electron localization function\cite{becke90,savin05} of $C2/m(2)$ at 2.7\,TPa,
where the weak covalent bonds along the molecular chains built up of H$_{3}$ clusters
are distinct.
It is important to point out that although $C2/m$ and $\beta$-Hg are
chained molecular phases, they are natural conductors. We do not need to perform
a band structure calculation to confirm this, since these phases have an odd number of electrons in their primitive cell,
therefore there always have a half-filled band.
Although $Pmmn$ has 6 electrons in its primitive cell, a DFT calculation with GGA
showed that it is metallic when pressure is beyond 600\,GPa.
This is different from lithium or sodium where intermediate compression induces localization of electrons
at interstitial regions and leads to metal-semiconductor-metal transition.\cite{lv11,ma09,marques11}

\begin{figure}
  \includegraphics*[width=3.0 in]{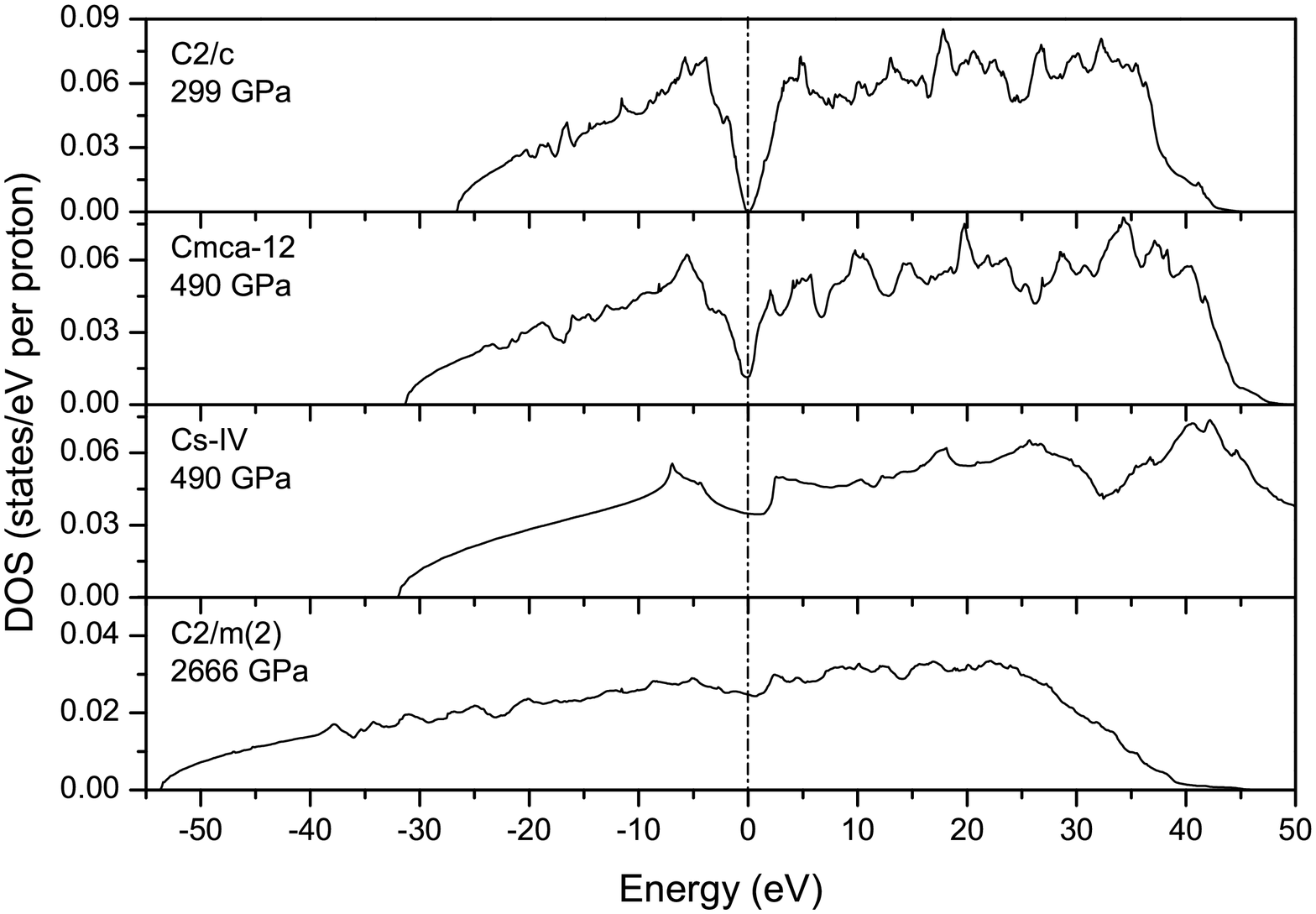}
  \caption{Electronic density of states of dense hydrogen at high pressures. The dash-dotted vertical line
  indicates the Fermi level. Notice that $C2/m(2)$ at a pressure of 2.7\,TPa has typical features
  of simple metals.}
  \label{fig:dos}
\end{figure}

The electronic density of states (DOS) of $C2/c$, $Cmca$-12, Cs-IV, and $C2/m(2)$ phases at selected pressures are shown
in Fig.\ref{fig:dos}, respectively. Being consistent with previous studies, the molecular hydrogen
at 299\,GPa does not show metallic characteristics, and there is an energy gap presented at the Fermi
level.\cite{pickard07} At a pressure of 490\,GPa in a diatomic molecular $Cmca$-12 phase, the gap already closes up and the material
becomes metallic. Here we didnot attempt to determine the precise insulator-metal transition pressure
in the molecular phase but instead focused on the general trend of the variation of DOS with pressure. It is interesting to
note that the DOS dips down at the Fermi level in $Cmca$-12, clearly showing that the closure of the
gap is due to band overlapping. This feature disappears after transition into the atomic Cs-IV at the
same pressure, it also absents from $C2/m(2)$ at about 2.7\,TPa in spite of covalent bonds
presenting in this meta-stable phase.
Overall, with increase of the pressure,
the localized covalent states in diatomic molecular phases become dispersive and
extend towards both the high and low energy
ends, which then closes up the gap by band overlapping at the Fermi level.
On the other hand, Cs-IV and $C2/m(2)$ show typical characteristics of simple metals in their DOS.
Especially, the DOS of $C2/m(2)$ has already been dispersed greatly by compression and becomes
flat and featureless over a wide range of energy. Other phases of dense hydrogen at ultra-high pressures
are similar and it is unnecessary to discuss them separately.

\subsection{Effects of zero-point motion of protons}

\subsubsection{Harmonic ZPE and its failure}
\begin{figure}
  \includegraphics*[width=3.0 in]{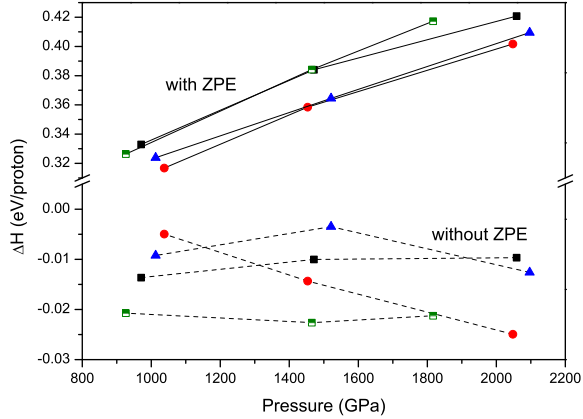}
  \caption{(color online) Change of the relative stability of meta-stable dense hydrogen due to contribution from
  harmonic ZPE, where the solid (dased) lines indicate with (without) ZPE,
  $\Delta H$ stands for the enthalpy difference with respect to a predesignated reference state, and
  the filled circles, squares, triangles, and half-filled squares denote $C2/m(2)$, $C2/m(1)$, $\beta$-Hg, and $Pmmn$
  phases, respectively.}
  \label{fig:H-ZPE}
\end{figure}

The contribution of zero-point vibrations of protons to
energy and enthalpy can be taken into account within quasi-harmonic approximation.
Since in a static lattice approximation our calculated groundstates and the relative stable order of high pressure phases
of dense hydrogen are almost the same as that reported in Ref.\onlinecite{mcmahon11}, and they had performed a detailed
analysis of the zero point energy (ZPE) in harmonic approximation, thus it is not necessary
to repeat the discussion. Here we only illustrate the magnitude and the possible
consequence of harmonic ZPE using meta-stable multi-atomic molecular phases as examples.
As Fig.\ref{fig:H-ZPE} shows, inclusion of harmonic ZPE slightly changes the relative stability of
these phases. The magnitude of ZPE is about 0.32\,eV per proton at 1\,TPa, and increases to
0.42\,eV per proton at 2\,TPa.
For highly compressed atomic phases of hydrogen, this treatment is inappropriate
because of the strong anharmonic effects.\cite{natoli93,natoli95,biermann98a,biermann98b}
However, for multi-atomic molecular phases, the anharmonic effects are also remarkable and
the shape of the potential well around the equilibrium position of each proton is far from being a quadratic form (see below),
which undermines the justification for harmonic approximation.
In this sense the relative stability of dense hydrogen cannot be faithfully determined by harmonic ZPE,
because the enthalpy difference between these phases is too small, and at the same time it is almost impossible
for the harmonic approximation to have a precision of within several percents in the case of dense hydrogen.

On the other hand, most of the energy barriers that separate different phases of dense hydrogen are much smaller than the magnitude
of harmonic ZPE. In this situation, the classical notation of
\emph{structural phase} might be ill-defined, because it is easy for
the system to overcome the barriers and travel freely from one \emph{phase} into another
driven by ZP motions.
Such kind of quantum fluctuation between structures invalidates not only harmonic approximation of lattice dynamics,
but also some restricted quantum Monte Carlo (QMC) treatments.\cite{natoli93,natoli95}
To tackle this problem quantitatively, a full quantum treatment of protons on the same footing as electrons is required,\cite{johnson98,ashcroft00}
which however, is
difficult within the DFT framework based on Born-Oppenheimer approximation.

\subsubsection{Anharmonic zero-point motion}

\begin{figure}
  \includegraphics*[width=3.0 in]{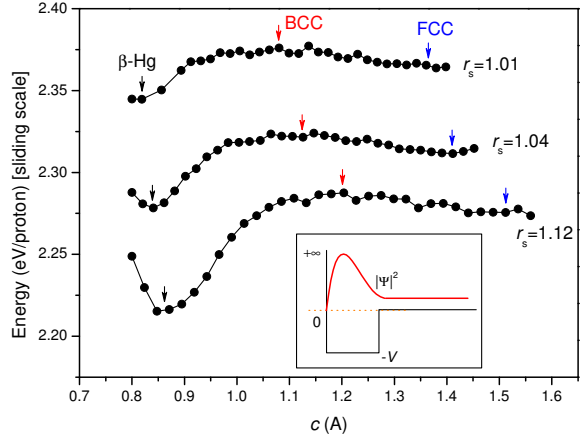}
  \caption{(color online) Energy variation with the change of the $c$ axial length of the $\beta$-Hg structure, which passes through FCC,
  BCC, and $\beta$-Hg successively. Inset: a simplified half-infinite potential well that models the energy surface, as well as the corresponding
  probability density profile of an effective particle moving in it.}
  \label{fig:E-BCC}
\end{figure}

Although in DFT it is difficult to carry out a quantitative analysis
of the anharmonic ZP effects if nuclei are treated as classical particles,
an insightful perception of the quantum motion of protons still can be obtained by
inspecting the energy surface or landscape closely.
Figure \ref{fig:E-BCC} shows a series of section of
the energy surface that cut along
a transition path from FCC to $\beta$-Hg at different densities.
That is to say, changing the lattice length $c$ of the $\beta$-Hg structure
while adjusting the $a$ and $b$ accordingly at a fixed density with $r_{s}$ equaling 1.12
($\sim$$1$\,TPa), 1.04 ($\sim$$1.5$\,TPa), and 1.01 ($\sim$$2$\,TPa), respectively.
Here the dimensionless parameter $r_{s}$ is defined as the radius of a sphere which encloses on the
average one electron in a unit of the Bohr radius.\cite{wigner34,carr61,jones96}
A value of $r_{s}=1$ corresponds to a physical compression ratio of about 30 for hydrogen.
The energy surface obtained in this way is approximation free, except those that already introduced in the standard DFT formalism.
Particles move on these surfaces at 0\,K as zero point motions.

It is easy to find from these surfaces that the harmonic approximation breaks down completely
for monatomic phases. There is even no potential well can be defined between BCC and FCC (and between BCC and $\beta$-Hg as well).
The flatness of the energy surface unveils an important origination of the unusually large
anharmonic effects in monatomic phases of hydrogen observed in QMC calculations (in addition to the light mass of protons):\cite{natoli93,natoli95,biermann98a,biermann98b}
having such a flat
energy surface, the crystal might melt even at zero Kelvin.

On the other hand, the harmonic ZP root mean square displacement (rMSD) gives a typical value of {0.2\,\AA} for $\beta$-Hg
phase.\cite{note_qha} It characterizes a typical ZP motion size for protons in a chained molecular phase.
Making use of this value and the potential well width given in Fig.\ref{fig:E-BCC},
we estimated that at a pressure of 1\,TPa the ZP motions of protons are confined within
the potential well. But with increasing of the pressure, ZP vibrational size
becomes comparable with the potential well width at 1.5\,TPa, and exceeds the latter at 2\,TPa.
Namely, beyond that pressure $\beta$-Hg might merge into FCC and BCC phases driven by ZP motions.
This picture can be understood by analogy with an effective particle that moves on a similar energy surface (a half-infinite potential well),
as shown in the inset of Fig.\ref{fig:E-BCC}. In that case a bound state (\emph{i.e.}, a well-defined crystal) exists only when
$8MVx^{2}\geq \pi^{2}\hbar^{2}$
where $x$($V$) is the potential width(depth) and $M$ is the particle mass.\cite{landauqm}
When no bound state is available, the particle moves around freely, but a phase cancellation
between the incident and reflected waves make it have a relative high probability
within the potential well, as the probability density profile in the inset of Fig.\ref{fig:E-BCC} illustrates.

Alternatively, this phenomenon can also be understood intuitively via the path integral formalism of quantum statistics theory.
The partition function of a system consisting of distinguishable particles (here hydrogen atoms)
can be written as\cite{ceperley95}
\begin{equation}
  Z=\int \rho(R,R;\beta)\,dR,
\end{equation}
with the diagonal density matrix given by
\begin{equation}
  \rho(R,R;\beta)=\rho^{0}(R,R;\beta)\left\langle\exp\left[-\int_{0}^{\beta}V(R_{t})\,dt\right]\right\rangle_{BW}.
  \label{eq:ruo}
\end{equation}
Here $\rho^{0}$ is the free particle density matrix, $\beta$ denotes $({k_{B}T})^{-1}$ where $T$ is the temperature,
and the potential contribution
(the $\langle\rangle_{BW}$ term) is given by averaging
over Brownian motions along closed paths that weighted by potential energy.
The total energy with ZP contribution included is thus
\begin{equation}
  E_{0}=-\lim_{\beta\rightarrow\infty}\frac{\partial\ln Z}{\partial\beta}.
\end{equation}
In cases where structural phases are separated by high enough barriers, the energy $E_{0}$ is
locally defined. Namely, all Brownian motions in Eq.(\ref{eq:ruo}) are effectively confined to a limited phase space which characterizes
the structure, and therefore one can compare the value of $E_{0}$ to determine the relative stability between different phases.

However, when the energy barrier is low or even no barrier at all, $E_{0}$ of one \emph{phase}
contains contribution of the Brownian paths winding across other \emph{phases}, and the definition
of the energy of that phase becomes meaningless. In this situation one could instead analyze the probability for a Brownian motion to fall
into a specific region in the phase space. Taking $\beta$-Hg for example, the probability for this phase to appear is given by
\begin{equation}
  \eta=\frac{\int_{\Omega_{m}}\rho(R,R;\beta)\,dR}{\int_{\Omega}\rho(R,R;\beta)\,dR},
\end{equation}
where $\Omega_{m}$ is the domain of the phase space defined by the structure of $\beta$-Hg, \emph{i.e.},
an analogue of the width of the potential trap as shown in Fig.\ref{fig:E-BCC}, and $\Omega$ is the whole phase space domain that is
accessible to all Brownian motions. Because the potential well is attractive, the paths that belong to/or pass
through the potential well ($\Omega_{m}$) always have a higher weight than others. Therefore $\eta$ has a non-zero
value at low enough temperatures. With the temperature decreases further, $\Omega$ gradually shrinks backwards to $\Omega_{m}$.
At zero Kelvin, if $\Omega$ reduces to $\Omega_{m}$ exactly, then the phase
of $\beta$-Hg is a well defined classical structure. Otherwise there are quantum fluctuations and $\beta$-Hg
exists only instantaneously with a probability of $\eta$.
Note this kind of quantum behavior that blurs the structural boundary of phases
had been noticed in path integral simulations,\cite{biermann98a,biermann98b}
which treated the quantum nature of protons.
Here we considered only meta-stable phases, similar argument (though not identical) can be applied to
the groundstate Cs-IV and its distortions.

\subsection{Energy barrier of phase transition}

\begin{figure}
  \includegraphics*[width=3.0 in]{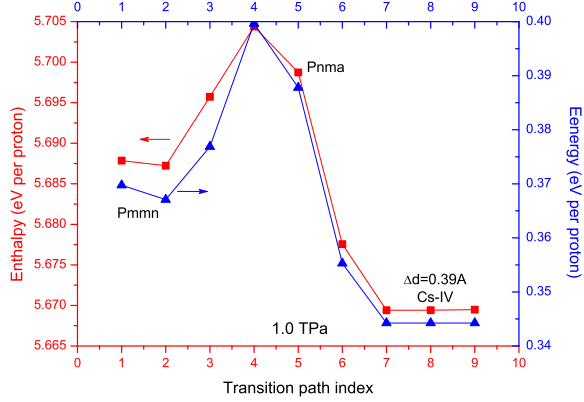}
  \caption{(color online) Energy and enthalpy variations along the phase transition path from $Pmmn$ to Cs-IV at 1\,TPa.
  Note that Cs-IV has a freedom to drift far away from its ideal position.}
  \label{fig:Pmmn2CsIV}
\end{figure}

From above discussion, we knew that the shape of the energy surface and the barrier that separates
different phases are crucial content to understand the high pressure behavior
of dense hydrogen comprehensively. The transition path shown in Fig.\ref{fig:E-BCC} is a special case,
which reveals that no energy barrier presents between chained $\beta$-Hg and cubic FCC
or BCC phases. Furthermore, the potential well around $\beta$-Hg becomes shallower and shallower with
increasing pressure, indicating the enhancement of the stability of high-symmetric phases and
the weakening of the tendency to form low coordinated structures.
It should be noted that this path is along a preassigned route. We can do this because the transition path is fixed
completely by symmetry and structure analysis. However, this is not always the case, and a
general technique such as NEB has to be employed in order to map out the transition path.

Figure \ref{fig:Pmmn2CsIV} illustrates the transition path from meta-stable $Pmmn$ to the groundstate
Cs-IV phase at 1\,TPa calculated with NEB. Both variations of energy and enthalpy are shown.
A distinct barrier (0.035\,eV/proton in enthalpy and 0.055\,eV/proton in energy from
the Cs-IV side) was clearly obtained. From the energy variation, we can see that both of the end
phases have distortions with a small energy change. In particular, the groundstate Cs-IV has various variants
which possess almost the same energy or enthalpy. The average drift distance of atoms between these variants
can be as large as $\triangle d=0.39$\,{\AA}, compared with the NN distance of 0.92\,{\AA} in this structure.
This implies that it might be quite common for broad and flat basins to present at some locally stable phases
(even the groundstate) of dense hydrogen, which was never noticed before. Furthermore, the flatness of the basins implies
that the ZPE should be much smaller than that predicted by harmonic approximation. In addition, since an enthalpy
barrier of 0.035\,eV/proton from Cs-IV to the first meta-stable phase $Pmmn$ corresponds to
a temperature scale of 400\,K, it is highly possible that dense hydrogen is in a solid state at this pressure
range at room temperature. Of course, considering that Cs-IV itself has many variants with negligible energy
change, it is also possible that it melts locally driven by ZP motions. If this is the case,
then a liquid-liquid transition could be expected when temperature is increased.

\begin{figure}
  \includegraphics*[width=3.0 in]{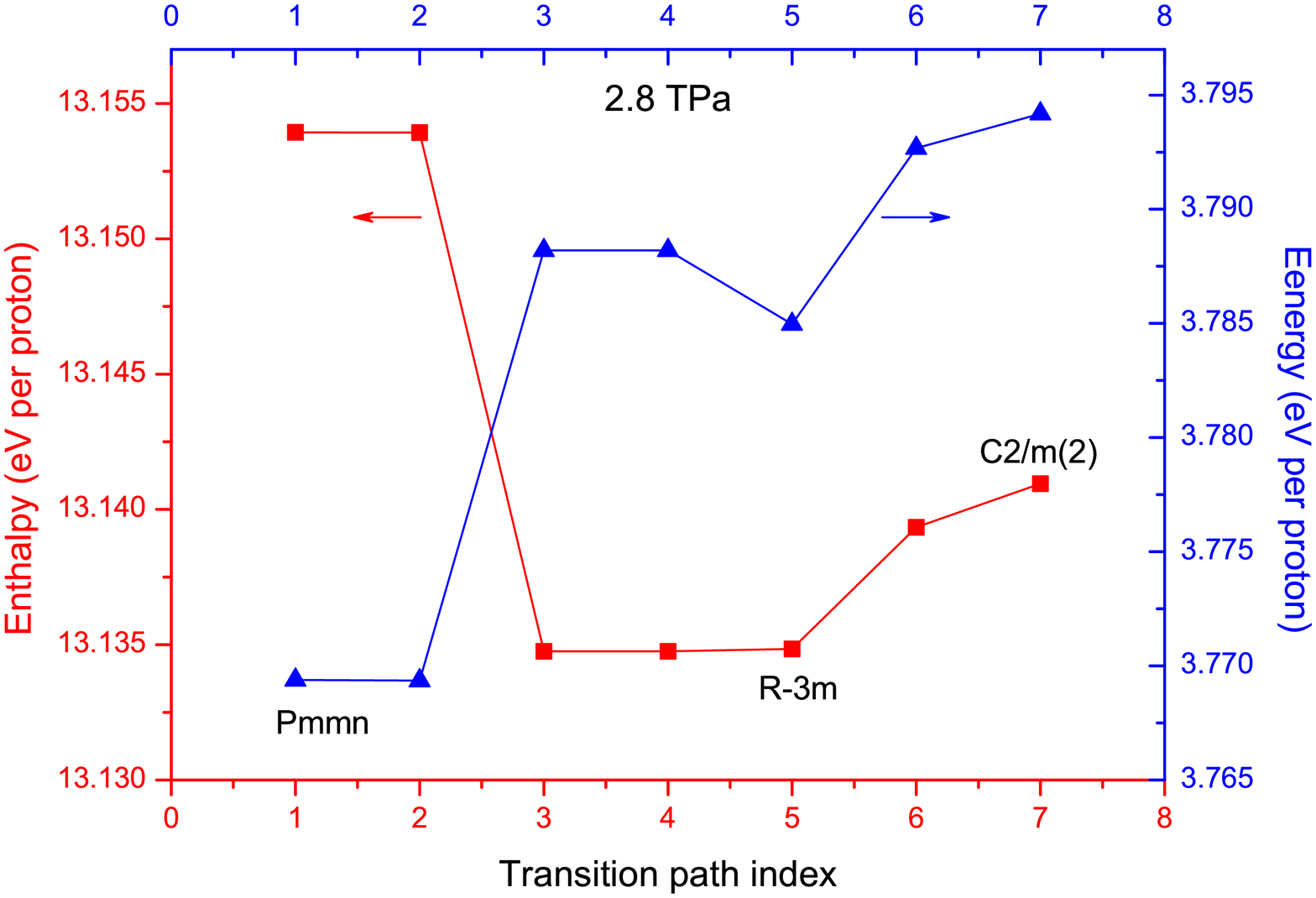}
  \caption{(color online) Energy and enthalpy variations along the phase transition path from $Pmmn$ to $C2/m(2)$ at 2.8\,TPa. Note that the energy
  variation does not comply with that of enthalpy, and a more stable phase of $R\overline{3}m$ is unveiled.}
  \label{fig:Pmmn2c2m2}
\end{figure}

Being analogous to Fig.\ref{fig:E-BCC}, a well-defined barrier is absent between the meta-stable $Pmmn$
and $C2/m(2)$ at 2.8\,TPa, as shown in Fig.\ref{fig:Pmmn2c2m2}. There are three noticeable
characteristics presented in this path: (i) a more stable $R\overline{3}m$ phase manifests itself in the transition path;
(ii) a flat energy (or enthalpy) surface appears again; and (iii) the energy variation
is inverse to that of enthalpy. The first point shows that it is possible to find out more stable
structures by investigating the transition path between high-lying meta-stable phases. This
is also helpful for understanding the physical mechanism of structure stability. The second
point reflects the frustration of different competing factors, and implies that it is
ineffective to optimize the structure by conventional relaxation algorithms because the forces become
too small to evolve the geometry within these flat regions. The final point clearly
implies that $C2/m(2)$ is stabilized by the term of $PV$, and should have great
imaginary phonon frequencies in harmonic approximation because the lattice dynamical matrix is evaluated on the
energy surface. However, such imaginary modes do not necessarily mean that the phase is locally unstable in thermodynamics
if one takes the $PV$ contribution into account.

\begin{figure}
  \includegraphics*[width=3.0 in]{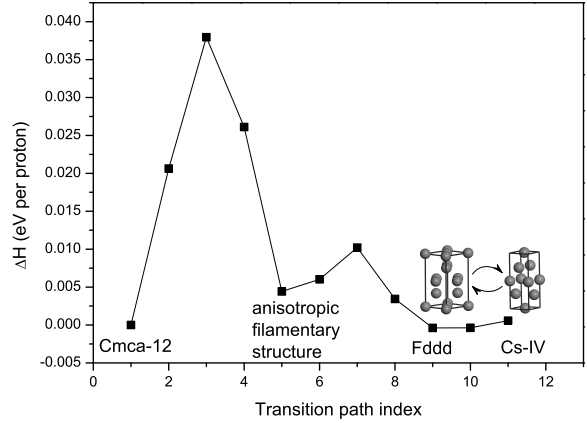}
  \caption{Enthalpy variation of dense hydrogen along the transition path from diatomic molecular $Cmca$-12 to atomic
  Cs-IV phase at 495\,GPa, where a low-symmetric filamentary phase and $Fddd$ phase (a distorted diamond structure) were observed.
}
  \label{fig:cmca12toCsIV}
\end{figure}

It is interesting to investigate the transition path from diatomic molecular $Cmca$-12 to atomic
Cs-IV at the dissociation pressure of 495\,GPa. As Fig.\ref{fig:cmca12toCsIV} shows, a high
enthalpy barrier of $\triangle H=0.038$\,eV/proton presents. This will introduce hysteresis and lift
the apparent dissociation pressure at low temperatures. Since the volume difference per atom between
$Cmca$-12 and Cs-IV is about $\triangle V=0.03$\,{\AA$^3$}, the possible pressure increase is thus
$\triangle P=\triangle H/\triangle V $$\approx 200$\,GPa to the first order of correction.\cite{geng07} That is to say, the DFT
predicted dissociation pressure of diatomic hydrogen should be at about 700\,GPa if taking the kinetic effect
of enthalpy barrier into account. Other factors that might also have some impacts include: (i) ZP motion
of protons, which favors atomic phases and therefore should decrease the dissociation pressure,\cite{natoli93,natoli95,straus77,mcmahon11}
but its precise value is hard to evaluate at present; (ii) intrinsic DFT error which favors
homogeneous distribution of electronic density, thus also underestimates the dissociation pressure about
50\,GPa;\cite{morales10pnas} (iii) other transition path with lower enthalpy barrier, we cannot exclude this possibility
completely because NEB is in fact a local optimization algorithm, and it explores a limited phase space
and thus depends on the initial guess of the transition path in some degree. Nevertheless,
after taking all of these factors into account the dissociation pressure should be less than 750\,GPa,
or 550\,GPa if no any hysteresis effect presents.

Another interesting phenomenon unveiled in Fig.\ref{fig:cmca12toCsIV} is that there is no energy
barrier between the degenerate groundstates of Cs-IV and $Fddd$. In other words, dense hydrogen
\emph{fluctuates} between these two structures driven by ZP motions. Note that although $Fddd$ can be viewed as a distortion
of Cs-IV structure, from a crystallographic point of view, it is in fact an orthorhombic variant of the
diamond phase. Namely, the quantum structural fluctuation in the groundstate of dense hydrogen is much more drastic
than what Fig.\ref{fig:Pmmn2CsIV} has been implied. Also an anisotropic low-symmetry filamentary phase
manifests itself on the transition path, which is very similar to the previously proposed
low-symmetric structures of dense hydrogen.\cite{biermann98a,biermann98b,brovman71,brovman72,kaxiras91} It is separated from other phases by an energy barrier
and might be meta-stable.

\subsection{Stability of dense hydrogen at zero pressure}

\begin{figure}
  \includegraphics*[width=3.0 in]{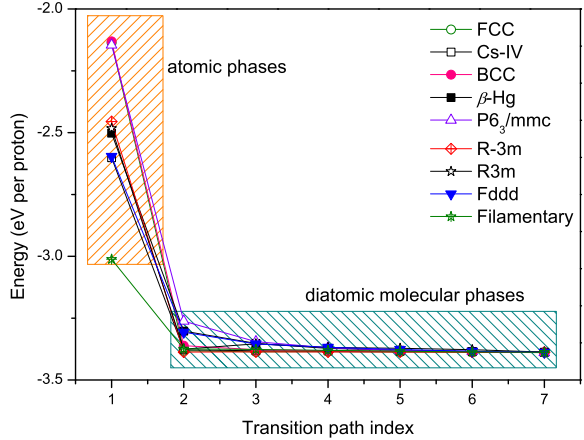}
  \caption{(color online) Energy variation of a series of dense hydrogen
  phases along the respective transition path to a diatomic molecular structure at zero pressure.
  No energy barrier can be detected. }
  \label{fig:denseH_0GPa}
\end{figure}

The stability of atomic phases of dense hydrogen at zero pressure is intriguing because
it is a possible room temperature superconductor.\cite{ashcroft68,zhang07,maksimov01,szczesniaka09,mcmahon11prb}
Although early primary perturbation calculations on
the structural energy of hydrogen within a static-lattice approximation
with the effective electron-ion interaction expanded to fourth order suggested that
a highly anisotropic structure might be stable,\cite{brovman71,brovman72}
later more careful analyses dismissed
this proposal and found that isotropic phases are more favored.\cite{straus77,natoli93,natoli95}
The energetic stability of these recently proposed new groundstates and low-lying meta-stable phases\cite{johnson00,pickard07,mcmahon11} at 0\,GPa,
however, has not been studied yet. Namely, whether these high pressure phases of hydrogen
is quenchable or not is still unknown.

To answer this question, we investigated the transition path and the associated energy barrier of these dense
phases to diatomic molecular structures at 0\,GPa with NEB calculations. The initial dense structures
were prepared by carefully relaxing the configurations from high pressures with constraints applied.
The corresponding diatomic molecular phases were then produced by distorting and relaxing
the respective dense structures. Typical results are illustrated in Fig.\ref{fig:denseH_0GPa}.
The conclusion is that all of the low-lying states of dense hydrogen known so far, including isotropic
atomic phases and low coordinated tri-atomic or chained molecular phases, have no
transition barriers can be detected along the transformation path.
In particular, the exotic filamentary structure observed in Fig.\ref{fig:cmca12toCsIV}
has no energy barrier, too.
That is to say, there is nothing that can prevent these
phases from spontaneously decaying to diatomic molecular structures. This fact that dense hydrogen might be \emph{unquenchable}
is a direct consequence of the strong tendency of hydrogen to pair at
low pressures ($Pmmn$ and $C2/m$ are also unstable at 0\,GPa,
which didnot include in Fig.\ref{fig:denseH_0GPa}).
This result is obtained with DFT in PBE approximation.
This exchange-correlation energy functional constructed on a local or semi-local approximation to the
homogeneous electron gas value overestimates the stability of a metallic phase slightly, but
cannot eliminate the whole energy barrier completely if there were one.
Also, an insightful analysis of the intrinsic error in DFT\cite{jones89,cohen08} implies that it seems unlikely
that an additional energy barrier can be predicted by an exact theory of DFT when
no any barrier can be detected with the current version of DFT. That is to say, even in a
level of the exact many-body quantum theory, dense hydrogen might still be unquenchable for these already known structures.

\subsection{Equation of state at zero Kelvin}

\begin{figure}
  \includegraphics*[width=3.0 in]{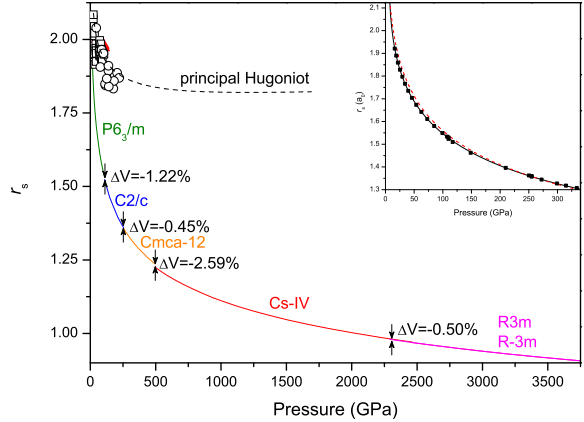}
  \caption{Calculated equation of state at 0\,K of dense hydrogen up to 3.5\,TPa
  compared with the extrapolation of the principal Hugoniot. Inset shows the comparison
  with the Vinet EOS (dashed line) that fitted to experimental data up to 119\,GPa at low pressures.}
  \label{fig:EOS}
\end{figure}

Equation of state (EOS: the pressure-volume relation here) is an important content to understand
the response of a material to compressions, which reflects not only the phase transitions driven
by pressure, but also the evolution of interactions with the change of coordination environment.
In experiment, x-ray diffraction measurements of the structure of single-crystal molecular
hydrogen had been performed at pressures up to 109\,GPa for H$_{2}$ and 119\,GPa for D$_{2}$.\cite{loubeyre96}
From these measurements a high-pressure EOS was deduced,\cite{loubeyre96} which is significantly more
compressible than that early constrained from lower-pressure data to 42\,GPa.\cite{mao94}
This deviation was interpreted as gradual effect of orientation of the H$_{2}$ molecular axis
within phase I.\cite{loubeyre96} At higher pressures, hydrogen transforms into atomic Cs-IV phase, where both
interatomic interactions and crystal structure are changed. Therefore there is enough
reason to suspect that this experimental EOS cannot be extrapolated to higher pressure range.

With the information of the calculated groundstate structures, the whole pressure-volume curve of dense hydrogen at
zero Kelvin was
calculated up to 3.5\,TPa with DFT. Figure \ref{fig:EOS} shows this curve by comparing with
the extrapolation of the principal Hugoniot. It is worthwhile to note that although there is a little
difference between $R3m$ and $R\overline{3}m$ in the enthalpy and it is hard to determine which one is the true
groundstate at ultra-high pressures, both phases have an almost identical $P$-$V$ relation. Similarly, $Fddd$ has the same
variation of density with pressure as that of Cs-IV and its distortions, and thus didnot include in Fig.\ref{fig:EOS}.
Along the groundstate line, the volume collapses due to phase transitions are small, it is about 2.59\% when dissociates to Cs-IV phase and 1.22\% between
$P6_{3}/m$ and $C2/c$. In other transitions it is just about 0.5\%.

\begin{figure}
  \includegraphics*[width=3.0 in]{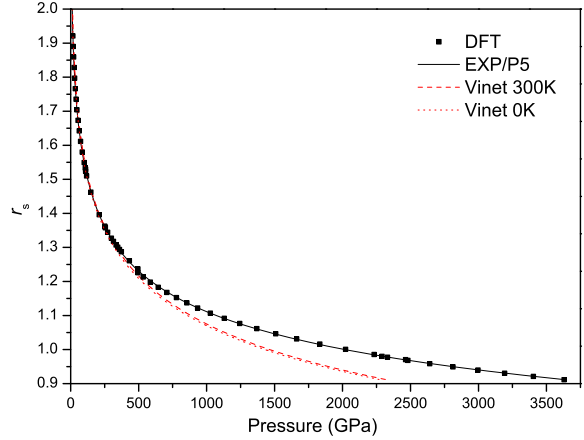}
  \caption{Comparison of the calculated pressure-volume curve with the Vinet EOS models
  that fitted to low-pressure data and a proposed EXP/P5 function
  within a pressure range up to 3.5\,TPa.}
  \label{fig:EOS2}
\end{figure}

The experimental data can be fitted to a Vinet function,\cite{loubeyre96} which is overall in good agreement with our DFT calculated
data below 300\,GPa, as the inset of Fig.\ref{fig:EOS} shows. However, there is a subtle discrepancy: the
DFT overestimates the compressibility slightly when between 50 and 150\,GPa. This might be due to the fact
that we didnot treat the quantum rotation of $H_{2}$ molecules explicitly. This degree of freedom
of motion should contribute energy and thus increase the internal pressure accordingly. At higher pressures,
however, extrapolation of the fitted Vinet function overestimates the compressibility significantly.
As Fig.\ref{fig:EOS2} illustrates, both of the Vinet function that fitted to original experimental data (Vinet 300\,K)
and that fitted to data reduced to 0\,K without ZP contribution (Vinet 0\,K) deviate
from the DFT results when beyond 500\,GPa, and the difference can reach as high as 36\% at a
density of $r_{s}=0.9$. Note that the curves of ``Vinet 300\,K'' and ``Vinet 0\,K'' are almost
identical within the studied pressure range, showing that both the ZP pressure and thermal pressure are
relatively small in comparison. Furthermore, since in DFT calculations we didnot take ZP contribution
into account and usually ZP motion contributes a positive pressure and thus should reduce the compressibility,
the DFT EOS shown in Fig.\ref{fig:EOS2} is thus an upper estimate of the compressibility
of dense hydrogen, and any model with a higher compressibility (\emph{e.g.},
the fitted Vinet EOS) is not allowed in physics.

If ignored the small volume collapses at the
phase transition points, the pressure-volume relation of the groundstates of dense hydrogen at 0\,K
calculated by DFT can be fitted excellently
by a function of
\begin{equation}
  P=\prod_{n=0}^{5}10^{A_{n}r_{s}^{n}}.
\end{equation}
We denote it as EXP/P5. When the pressure $P$ is in a unit of GPa, the parameters are as follows:
$A_{0}=1.0683$, $A_{1}=19.1824$, $A_{2}=-36.3776$, $A_{3}=28.5165$, $A_{4}=-10.6068$, and $A_{5}=1.5224$.
It should be noted that EXP/P5 function not only faithfully represents the overall variation of the EOS of dense
hydrogen over a broad range of pressure up to 3.5\,TPa, it also reproduces the low pressure data
accurately, as demonstrated in the inset of Fig.\ref{fig:EOS}. With these properties, this function should
have a good applicability for extrapolating the EOS of hydrogen to higher pressures beyond several TPa.

\section{Conclusion}
\label{sec:sum}

High pressure phases of dense hydrogen have been extensively searched up to 24 atoms per unit cell and
3.5\,TPa in pressure with density functional theory calculations.
The results confirmed the previous conclusion that diatomic hydrogen dissociates into atomic Cs-IV
at 495\,GPa and then transforms to $R3m$ or $R\overline{3}m$ at about 2.3\,TPa.
Exotic high pressure behaviors were also discovered that show dense hydrogen having numerous
isomorphs with comparable enthalpy, and frustrated competitions among them lead to broad and flat basins on the energy surface.
In particular, calculations showed that there is even no energy barrier to separate
crystallographic different phases in some cases. Furthermore, the atomic groundstate Cs-IV has a freedom
to distort and is degenerate with orthorhombic $Fddd$, which fluctuates forth and back to
Cs-IV driven by ZP motions of protons.

Within a wide range of pressure beyond 500\,GPa, the first meta-stable structure is a tri-atomic $Pmmn$,
therefore the stability of the closely related multi-atomic molecular phases was also analyzed. Electronic structure calculations
indicated that they are metallic in nature, but with weak covalent bonds presented. The occurrence of these
meta-stable phases is a direct consequence of the competition between the tendency to overlap
the electron orbitals among neighboring atoms and the pressure-induced dissociation of molecular hydrogen.
The structural relationship between these exotic phases was investigated for a better understanding
of the structural behavior of dense hydrogen.

In addition to static-lattice calculations, ZP vibrations of lattice were also computed in the harmonic approximation and the magnitude of its
contribution was estimated. Inspection of the energy surface (via transition path) dismissed the validity
of this level approximation, and the anharmonicity of ZP motions was demonstrated by studying
the variation of the energy surface with pressure, which not only reveals the weakening of the tendency
to form covalent bonds under compression, but also illustrates the importance of quantum fluctuations
in structure of dense hydrogen at high pressures.

The general transition path and the associated energy barrier between different phases were calculated with NEB technique, which suggested that
the dissociation pressure of diatomic molecular hydrogen might be deferred to at about 750\,GPa due to a hysteresis effect
of the energy barrier at low temperatures. The enthalpy barrier between the groundstate Cs-IV and the first meta-stable $Pmmn$
phase at 1\,TPa implies atomic hydrogen might be in a solid state at zero Kelvin, but it depends
on the exact contribution of anharmonic ZP vibrations of protons. The calculated transition path also
suggested that there are phases in dense hydrogen that are stabilized purely by $PV$ term, which is
unusual in the common sense. The meta-stability of dense phases of hydrogen at zero pressure were extensively
studied. No energy barrier was detected between them and the diatomic molecular phase, implying dense hydrogen might be unquenchable.

The equation of state (\emph{i.e.}, the pressure-volume relation) of dense hydrogen up to 3.5\,TPa was investigated.
A new EOS function, namely EXP/P5, was proposed, which can represent the calculated DFT data over the whole
studied pressure range excellently.
Since no thermal pressure and ZP contribution were included, this EOS is an upper estimation
of the true compressibility, provided that the current understanding of the groundstate structure of dense hydrogen is correct.
At low pressures, this EOS is in good agreement with the experimental data measured by single
crystal x-ray technique. Extrapolating the Vinet EOS model that fitted to these measured low-pressure data,
however, drastically overestimates the compressibility when beyond 500\,GPa, due to inappropriate
counting of the interatomic repulsion under extreme compressions in this model.
By the way, as isotopes of hydrogen, deuterium and tritium have almost the same electronic behavior.
The main difference is that they have a heavier ionic mass, and thus the ZP effects should be
less significant. But the overall picture of the physics is the same.

\begin{appendix}
\section{Justification to the density functional theory}
The sensitivity of the calculated results to the choice of exchange-correlation
density functional had been checked in Ref.\onlinecite{pickard07}. We repeated the checking process
and obtained similar conclusions. Furthermore, the validity of the density functional
theory to ultra-high pressure physics can be understood easily. We know that the apparent
failure of all local density approximation (LDA) based functionals relates directly to an abrupt
variation of charge density. Hydrostatic compression reduces interatomic distance,
which increases interactions between electrons and nearby nuclei.
Covalent bonds are thus weakened and electronic wave-function spreads out due
to environmental changes in atomic coordination usually.
The direct consequence is thus a more smooth spatial distribution
of the charge density, and a better performance of LDA-based
functionals with an increase of the pressure (note that the peculiar phenomenon of pressure-induced localization
of electrons observed in lithium\cite{lv11,marques11} and sodium\cite{ma09} is absent from dense hydrogen).
In addition, the \emph{s} orbital electrons, the only one that
is relevant in hydrogen, are always well described by LDA even at ambient conditions.

On the other hand, although a reliable convergence of the DFT total energy to within 3\,\emph{m}eV per proton has been
achieved, a concern that whether DFT can be applied to dense hydrogen beyond insulator-metal
transition might still exist because it is well known that DFT underestimates band gap and then
the transition pressure. It is a severe problem for electronic structure properties, but
has limit impacts on the total energy and geometry features. Theoretically DFT only ensures the correct
charge density and the total energy, rather than the quasi-particle levels that are introduced via Kohn-Sham ansatz.
For example in an extreme case where the exact exchange-correlation functional were available, DFT would become rigorous
and predict an exact charge density and total energy, whereas the band structure still cannot
be generally guaranteed theoretically.\cite{jones89,savin98} On the other hand, there are many examples in the applications of DFT to
condensed matters where the band structure given by DFT is wrong but the energetics and atomic structure are still within
an acceptable precision. Therefore the underestimation of the band gap of dense hydrogen by DFT
does not present as a serious issue to what we concerned here. The intrinsic error in the
total energy and structural features comes from the GGA approximation of the energy functional,
and should be similar for insulator and metallic phases as long as the spatial variation of the charge
density is similar.

It also needs to point out that near the transition point from insulator to metallic state,
where the band gap begins closing up and electrons jump abruptly from localized to delocalized
orbitals, DFT might underestimate the transition pressure,\cite{morales10pnas} presumably due to self-interaction errors
in exchange-correlation functional. But far away from this transition region, DFT
works well.\cite{morales10pnas,morales10pre} Since exact-exchange calculations predicted a metallization pressure
of hydrogen at 400\,GPa,\cite{stadele00}
much lower than the pressure range interested here, we estimate that the influence on the total energy and
structure features due to the errors
in delocalization of electrons occurred at low pressures should be small.

\section{Justification to the pseudopotential}

\begin{figure}
  \includegraphics*[width=3.0 in]{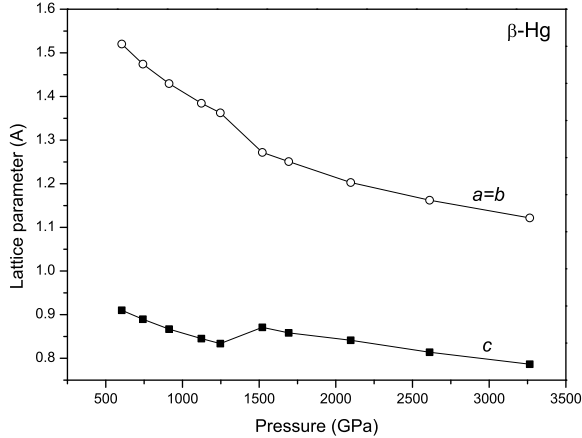}
  \caption{Variation of the tetragonal lattice parameters of $\beta$-Hg
  under compression, where the shortest lattice vector $c$, which indicates the bond length
  of the molecular chain, is always longer than the ambient H$_{2}$ value of $0.74$\,\AA.
  }
  \label{fig:Hg-abc}
\end{figure}

We used a PAW pseudopotential that is harder than the standard one, which is specially designed to account
for the possible short interatomic distance under ultra-high compressions. The performance
of this potential was checked by comparing
the bond length and binding energy of H$_{2}$ dimer with those of other potentials and all-electron
calculations. The low-lying structures and the corresponding transition pressures
below 400\,GPa as reported in Ref.\onlinecite{pickard07} were also perfectly reproduced.
This verified that the hard potential works correctly. Its
applicability to higher pressure range can be generally guaranteed as long as the shortest interatomic
distance is greater than twofold of the outmost cutoff radius $r_{m}$ of the potential,
which defines the atomic spheres (augmentation regions) where the pseudo wave function takes effects.
This precondition ensures the correct
wave-function shape near the atomic spheres. For comparison, $r_{m}$=0.42\,{\AA} for
the hard potential and $r_{m}$=0.58\,{\AA} for
the standard potential of hydrogen. In our studied pressure range here, no low-lying structure has a shortest
interatomic distance less than the ambient hydrogen molecule bond length of 0.74\,{\AA}.
For example, Fig.\ref{fig:Hg-abc} shows the variation of the lattice parameters of $\beta$-Hg with pressure,
where $c$ indicates the shortest bond length and is always greater than 0.74\,{\AA}.
The overlapping between atomic spheres of dense hydrogen is less than that of H$_{2}$ molecule at ambient
conditions. Therefore the application of the PAW potential of hydrogen to high pressures does not
present as a difficult issue.
It is worthwhile to point out that for ambient H$_{2}$, there is a little overlap of the
pseudopotential regions on nearest neighboring hydrogen atoms. In principle these spheres
shouldnot overlap, but a bit of overlap/softness-of-the-potential tradeoff is usually acceptable.
Especially when spherical Bessel functions were used to construct the PAW potential, as
implemented in VASP code, a large overlap between the atomic spheres is allowed.\cite{kresse99}

\begin{figure}
  \includegraphics*[width=3.0 in]{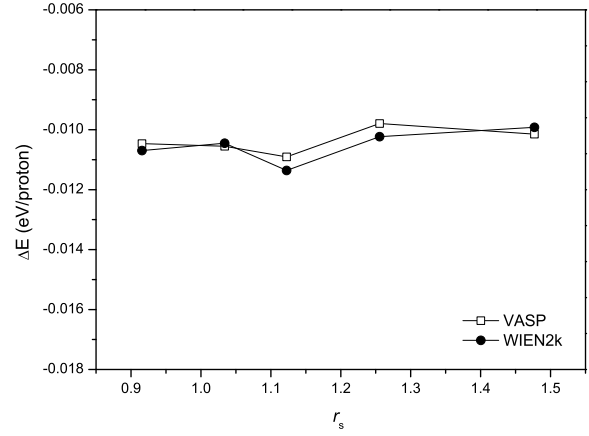}
  \caption{Comparison of the energy difference between FCC and BCC phases as a function of density
  calculated using PAW pseudopotential method with VASP from that of all-electron results
  of WIEN2k.
  }
  \label{fig:FCC-BCC}
\end{figure}

An overall examination of the validity of the PAW potential is illustrated in Fig.\ref{fig:FCC-BCC},
where the VASP result is compared directly with that of the all-electron method calculated with WIEN2k code.\cite{wien2k}
Both calculations used the same PBE exchange-correlation energy functional. We can see from
this figure that within
a wide range of pressure up to 4\,TPa ($r_{s}\approx 0.9$), both methods give almost the same energy difference between high-symmetric
BCC and FCC structures (the maximum deviation is less than 0.4\,\emph{m}eV per proton), reflecting the fact that the error introduced by PAW potential is insignificant.

\section{Structure of some meta-stable phases}
The structural information (including both lattice parameters and atomic coordinates) of
some meta-stable multi-atomic molecular phases and the degenerate groundstate $Fddd$ at
selected pressures are
listed in table \ref{tab:c2m2}.

\begin{table}[]
\caption{\label{tab:c2m2} Structure of some low-lying phases. Only the fractional coordinates of symmetry inequivalent
atoms are reported. The numbers of atoms in a primitive cell are given. $Immm$ is a coarse but robust (with greater tolerance) orthorhombic representation
of $C2/m$ structure. For $\beta$-Hg, a tetragonal distortion of the BCC cubic cell is adopted.}
\begin{ruledtabular}
\begin{tabular}{l c c c c c c c c } 
  space group & pressure & \multicolumn{3}{c}{lattice parameter} & & \multicolumn{3}{c}{atomic coordinates} \\
  \cline{3-5}\cline{7-9}
     $\sharp$ atoms& (GPa)  & \multicolumn{3}{c}{(\AA, $^{\circ}$)} &  & \multicolumn{3}{c}{(fractional)} \\
\hline
  $C2/m(1)$  & 650 &$a$=3.178 & $b$=1.203 & $c$=2.395   & H1 & 0.9999 & 0.0000 &  0.8300 \\
  3     &   & \multicolumn{2}{c}{$\alpha$=$\gamma$=90.00} &  $\beta$=137.63  & H2 & 0.5000 & 0.0000 &0.5000 \\
\\
  $Immm$ &    & $a$=2.142   & $b$=1.203   & $c$=2.395           &H1 & 0.5000 & 0.0000 & 0.6766   \\
  3 &         & \multicolumn{3}{c}{$\alpha$=$\beta$=$\gamma$=90.00} &H2 & 0.0000 & 0.0000 & 0.5000  \\
\hline
  $C2/m(2)$  & 1000 &$a$=3.180 & $b$=1.101 & $c$=2.711   & H1 & 0.0006 & 0.0000 &  0.3376  \\
  3     &   & \multicolumn{2}{c}{$\alpha$=$\gamma$=90.00} &  $\beta$=147.51  & H2 & 0.5000 & 0.0000 &0.0000 \\
\\
  $Immm$ &    & $a$=1.708   & $b$=1.101   & $c$=2.711           &H1 & 0.0000 & 0.0000 & 0.0000  \\
  3 &         & \multicolumn{3}{c}{$\alpha$=$\beta$=$\gamma$=90.00} &H2 & 0.5000 & 0.0000 & 0.1674 \\
\hline
  $Pmmn$  & 1104 &$a$=2.349 &$b$=1.110 & $c$=1.877                    & H1 & 0.8170& 0.5000& 0.6886  \\
  6     &   & \multicolumn{3}{c}{$\alpha$=$\beta$=$\gamma$=90.00}     & H2 & 0.0000& 0.5000& 0.1087\\
\hline
  $\beta$-Hg ($I4/mmm$)  & 1012 &\multicolumn{2}{c}{$a$=$b$=1.408} & $c$=0.856   & H1 & 0.0000 & 0.0000 &  0.0000  \\
  1     &   & \multicolumn{3}{c}{$\alpha$=$\beta$=$\gamma$=90.00}     &  &  &  & \\
\hline
  $Fddd$  & 533 &$a$=3.042 &$b$=1.832 & $c$=1.596                  & H1 & 0.7500 & 0.7500 &0.2500  \\
  2     &   & \multicolumn{3}{c}{$\alpha$=$\beta$=$\gamma$=90.00}     &  &  &  & \\
\end{tabular}
\end{ruledtabular}
\end{table}

\end{appendix}


\begin{thebibliography}{99}
\bibitem{guillot99a} T. Guillot, Science \textbf{286}, 72 (1999).
\bibitem{guillot99b} T. Guillot, Planet. Space Sci. \textbf{47}, 1183 (1999).
\bibitem{bonev04} S. A. Bonev, E. Schwegler, T. Ogitsu, and G. Galli, Nature \textbf{431}, 669 (2004).
\bibitem{deemyad08} S. Deemyad and I. F. Silvera, Phys. Rev. Lett. \textbf{100}, 155701 (2008).
\bibitem{morales10pnas} M. A. Morales, C. Pierleoni, E. Schwegler, and D. M. Ceperley, Proc. Natl. Acad. Sci. U.S.A. \textbf{107}, 12799 (2010).
\bibitem{lorenzen10} W. Lorenzen, B. Holst, and R. Redmer, Phys. Rev. B \textbf{82}, 195107 (2010).
\bibitem{mao94} H. K. Mao and R. J. Hemley, Rev. Mod. Phys. \textbf{66}, 671 (1994) (and references therein).
\bibitem{eggert91} J. H. Eggert, F. Moshary, W. J. Evans, H. E. Lorenzana, K. A. Goettel, I. F. Silvera, and
W. C. Moss, Phys. Rev. Lett. \textbf{66}, 193 (1991).
\bibitem{narayana98} C. Narayana, H. Luo, J. Orloff, and A. L. Ruoff, Nature \textbf{393}, 46 (1998).
\bibitem{loubeyre02} P. Loubeyre, F. Occelli, and R. Le Toullec, Nature \textbf{416}, 613 (2002).
\bibitem{bafile08} U. Bafile, F. Becherini, D. Colognesi, and M. Zoppi, Phys. Rev. B \textbf{77}, 224302 (2008).

\bibitem{wigner35} E. Wigner and H. B. Huntington, J. Chem. Phys. \textbf{3}, 764 (1935).
\bibitem{friedli77} C. Friedli and N. W. Ashcroft, Phys. Rev. B \textbf{16}, 662 (1977).
\bibitem{barbee89} T. W. Barbee, III, A. Garcia, M. L. Cohen, and J. L. Martins, Phys. Rev. Lett. \textbf{62}, 1150 (1989).
\bibitem{chacham91} H. Chacham and S. G. Louie, Phys. Rev. Lett. \textbf{66}, 64 (1991).
\bibitem{johnson00} K. A. Johnson and N. W. Ashcroft, Nature \textbf{403}, 632 (2000).
\bibitem{pickard07} C. J. Pickard and R. J. Needs, Nature Phys. \textbf{3}, 473 (2007).
\bibitem{tse08} J. S. Tse, D. D. Klug, Y. S. Yao, Y. L. Page, and J. R. Rodgers,
Solid State Commun. \textbf{145}, 5 (2008).
\bibitem{mcmahon11} J. M. McMahon and D. M. Ceperley, Phys. Rev. Lett. \textbf{106}, 165302 (2011).

\bibitem{wigner34} E. Wigner, Phys. Rev. \textbf{46}, 1002 (1934).
\bibitem{carr61} W. J. Carr, Jr., Phys. Rev. \textbf{122}, 1437 (1961).
\bibitem{jones96} M. D. Jones and D. M. Ceperley, Phys. Rev. Lett. \textbf{76}, 4572 (1996).

\bibitem{geng07}  H. Y. Geng, Y. Chen, Y. Kaneta, and M. Kinoshita, Phys. Rev. B \textbf{75}, 054111 (2007).
\bibitem{loubeyre96} P. Loubeyre, R. LeToullec, D. Hausermann, M. Hanfland, R. J. Hemley, H. K. Mao, and L. W. Finger, Nature \textbf{383}, 702 (1996).

\bibitem{pickard06} C. J. Pickard and R. J. Needs, Phys. Rev. Lett. \textbf{97}, 045504 (2006).
\bibitem{rousseau11} B. Rousseau, Y. Xie, Y. Ma, and A. Bergara, Eur. Phys. J. B \textbf{81}, 1 (2011).
\bibitem{lv11} J. Lv, Y. Wang, L. Zhu, and Y. Ma, Phys. Rev. Lett. \textbf{106}, 015503 (2011).
\bibitem{yao09} Y. Yao, J. S. Tse, and D. D. Klug, Phys. Rev. Lett. \textbf{102}, 115503 (2009).
\bibitem{wang10} Y. Wang, J. Lv, L. Zhu, and Y. Ma, Phys. Rev. B \textbf{82}, 094116 (2010).

\bibitem{kresse96} G. Kresse and J. Furthm{\"u}ller, Phys. Rev. B \textbf{54}, 11169 (1996).
\bibitem{blochl94} P. E. Bl{\"o}chl, Phys. Rev. B \textbf{50}, 17953 (1994).
\bibitem{kresse99} G. Kresse and D. Joubert, Phys. Rev. B \textbf{59}, 1758 (1999).
\bibitem{pbe96} J. P. Perdew, K. Burke, and M. Ernzerhof, Phys. Rev. Lett. \textbf{77}, 3865 (1996).
\bibitem{mcmahon06} M. I. McMahon and R. J. Nelmes, Chem. Soc. Rev. \textbf{35}, 943 (2006).

\bibitem{mills95} G. Mills, H. Jonsson, and G. K. Schenter, Surf. Sci. \textbf{324}, 305 (1995).
\bibitem{phon} D. Alf\`{e}, Comp. Phys. Comm. \textbf{180}, 2622 (2009).

\bibitem{note_ref} This state does not correspond to a physical one. It is artificially defined
by the enthalpy only for the purpose to unveil the subtle enthalpy difference among competing phases.
\bibitem{peierls55} R. E. Peierls, \emph{Quantum Theory of Solids} (Oxford University Press, London, 1955).

\bibitem{becke90} A. D. Becke and K. E. Edgecombe, J. Chem. Phys. \textbf{92}, 5397 (1990).
\bibitem{savin05} A. Savin, J. Mol. Struct.: THEOCHEM \textbf{727}, 127 (2005).

\bibitem{marques11} M. Marques, M. I. McMahon, E. Gregoryanz, M. Hanfland, C. L. Guillaume, C. J. Pickard, G. J. Ackland, and R. J. Nelmes, Phys. Rev. Lett. \textbf{106}, 095502 (2011).
\bibitem{ma09} Y. Ma, M. Eremets, A. Oganov, Y. Xie, I. Trojan, S. Medvedev, A. Lyakhov, M. Valle, and V. Prakapenka, Nature (London) \textbf{458}, 182 (2009).

\bibitem{natoli93} V. Natoli, R. M. Martin, and D. M. Ceperley, Phys. Rev. Lett. \textbf{70}, 1952 (1993).
\bibitem{natoli95} V. Natoli, R. M. Martin, and D. M. Ceperley, Phys. Rev. Lett. \textbf{74}, 1601 (1995).
\bibitem{biermann98a} S. Biermann, D. Hohl, and D. Marx, Solid State Commun. \textbf{108}, 337 (1998).
\bibitem{biermann98b} S. Biermann, D. Hohl, and D. Marx, J. Low Temp. Phys. \textbf{110}, 97 (1998).

\bibitem{johnson98} K. Johnson and N. W. Ashcroft, J. Phys.: Condens. Matter \textbf{10}, 11135 (1998).
\bibitem{ashcroft00} N. W. Ashcroft, J. Phys.: Condens. Matter \textbf{12}, A129 (2000).
\bibitem{note_qha} This value is comparable with the root-mean-square radius of gyration of a nucleus in \emph{ab initio} path integral simulations that
have taken the quantum effects of protons into account fully, see Refs.\onlinecite{biermann98a,biermann98b}.

\bibitem{landauqm} L. D. Landau and E. M. Lifshitz, \emph{Quantum Mechanics (Non-relativistic Theory)}, 3rd edition (Elsevier, Singapore, 1965).
\bibitem{ceperley95} D. M. Ceperley, Rev. Mod. Phys. \textbf{67}, 279 (1995).
\bibitem{straus77} D. M. Straus and N. W. Ashcroft, Phys. Rev. Lett. \textbf{38}, 415 (1977).

\bibitem{brovman71} E. G. Brovman, Y. Kagan, and A. Kholas, Zh. Eksp. Theor. Fiz. \textbf{61}, 2429 (1971) [Sov. Phys. JETP \textbf{34}, 1300 (1972)].
\bibitem{brovman72} E. G. Brovman, Y. Kagan, and A. Kholas, Zh. Eksp. Theor. Fiz. \textbf{62}, 1492 (1972) [Sov. Phys. JETP \textbf{35}, 783 (1972)].
\bibitem{kaxiras91} E. Kaxiras, J. Broughton, and R. J. Hemely, Phys. Rev. Lett. \textbf{67}, 1138 (1991).

\bibitem{ashcroft68} N. W. Ashcroft, Phys. Rev. Lett. \textbf{21}, 1748 (1968).
\bibitem{zhang07} L. J. Zhang, Y. L. Niu, Q. Li, T. Cui, Y. Wang, Y. M. Ma, Z. He, and G. T. Zou,
Solid State Commun. \textbf{141}, 610 (2007).
\bibitem{maksimov01} E. G. Maksimov and D. Y. Savrasov, Solid State Commun. \textbf{119}, 569 (2001).
\bibitem{szczesniaka09} R. Szczesniaka and M. W. Jarosik, Solid State Commun. \textbf{149}, 2053 (2009).
\bibitem{mcmahon11prb} J. M. McMahon and D. M. Ceperley, Phys. Rev. B \textbf{84}, 144515 (2011).

\bibitem{jones89} R. O. Jones and O. Gunnarsson, Rev. Mod. Phys. \textbf{61}, 689 (1989).
\bibitem{cohen08} A. J. Cohen, P. Mori-Sanchez, and W. Yang, Science \textbf{321}, 792 (2008).
\bibitem{savin98} A. Savin, C. J. Umrigar, and X. Gonze, Chem. Phys. Lett. \textbf{288}, 391 (1998).
\bibitem{morales10pre} M. A. Morales, C. Pierleoni, and D. M. Ceperley, Phys. Rev. E \textbf{81}, 021202 (2010).
\bibitem{stadele00} M. St{\"a}dele and R. M. Martin, Phys. Rev. Lett. \textbf{84}, 6070 (2000).

\bibitem{wien2k} P. Blaha, K. Schwarz, G. K. H. Madsen, D. Kvasnicka, and J. Luitz, \textbf{WIEN2k}, An Augmented
Plane Wave + Local Orbitals Program for Calculating Crystal Properties
(Karlheinz Schwarz, Techn. Universit{\"a}t Wien, Austria), 2001. ISBN 3-9501031-1-2


\end{thebibliography}
\end{document}